\documentclass[aps,prd,nofootinbib,amsmath,twocolumn]%{revtex4}
{revtex4-1}
\usepackage{amssymb,esvect,amsmath,graphicx,latexsym,amsthm,slashed,eso-pic,hyperref}
\usepackage[final]{pdfpages}
\newcommand{\be}{\begin{equation}}
\newcommand{\ee}{\end{equation}}

\newcommand{\GeV}{\text{ GeV}}
\newcommand{\MeV}{\text{ MeV}}
\newcommand{\keV}{\text{ keV}}
\newcommand{\eV}{\text{ eV}}
\newcommand{\rhodm}{\rho_{_\text{DM}}}

\def\lsim{\mathrel{\raise.3ex\hbox{$<$\kern-.75em\lower1ex\hbox{$\sim$}}}}
\def\gsim{\mathrel{\raise.3ex\hbox{$>$\kern-.75em\lower1ex\hbox{$\sim$}}}}

\newcommand{\order}[1]{\mathcal{O}{(#1)}}

\begin{document}

\hspace{13cm}

\title{Neutrino Oscillations as a Probe of Light Scalar Dark Matter}
\author{Asher Berlin,$^1$}

\affiliation{$^1$ Department of Physics, Enrico Fermi Institute, University of Chicago, Chicago, IL}

\date{\today}

\begin{abstract}
We consider a class of models involving interactions between ultra-light scalar dark matter and Standard Model neutrinos. Such couplings modify the neutrino mass splittings and mixing angles to include additional components that vary in time periodically with a frequency and amplitude set by the mass and energy density of the dark matter. Null results from recent searches for anomalous periodicities in the solar neutrino flux strongly constrain the dark matter-neutrino coupling to be orders of magnitude below current and projected limits derived from observations of the cosmic microwave background. 
\end{abstract}

\maketitle

Over the past several decades, the Weakly Interacting Massive Particle (WIMP) paradigm has motivated a plethora of dedicated searches for dark matter (DM) particles with weak-scale masses and couplings to the Standard Model (SM). Current direct detection experiments such as LUX~\cite{Akerib:2015rjg}, XENON100~\cite{Aprile:2012nq}, PandaX~\cite{Tan:2016zwf}, and SuperCDMS~\cite{Agnese:2013jaa} are sensitive to $\sim$~keV nuclear recoils, corresponding to the kinetic energy of DM with a mass of $\sim$~GeV or greater. In the near future, ton-scale detectors will continue to probe WIMPs of similar mass, but with increasingly smaller scattering cross sections with nuclei~\cite{Aprile:2015uzo,Akerib:2015cja}. DM with MeV - GeV scale mass is also well-motivated and is embodied in many models~\cite{Izaguirre:2015yja,Kaplan:2009ag,Boehm:2003hm,Pospelov:2007mp,Hooper:2008im,Feng:2008ya,Morrissey:2009ur,Hochberg:2014dra,Hochberg:2014kqa}. As a result, new strategies have been proposed to search for DM as light as the warm thermal DM limit, $\sim$~keV~\cite{Schutz:2016tid,Hochberg:2015fth,Hochberg:2015pha}.

However, the mass range of keV$-$TeV only spans a tiny fraction of the entire plausible DM landscape, which extends from $\sim 10^{-22} \eV$ to astrophysical scales~\cite{Khlopov:1985jw,Press:1989id,Sahni:1999qe,Hu:2000ke,Marsh:2015xka,Hlozek:2014lca,Bozek:2014uqa,Sarkar:2015dib,Schive:2014dra,Hlozek:2016lzm,Amendola:2005ad}. Furthermore, for a local density of $\rhodm \simeq 0.4 \text{ GeV} \text{ cm}^{-3}$, DM has a large phase space occupancy when its mass is below $\sim 0.1$ eV, and instead of a \emph{particle}, behaves more like an oscillating classical \emph{field}. The best known example of DM in this ``field-like" mass regime is the QCD axion~\cite{Peccei:1977hh,Weinberg:1977ma,Wilczek:1977pj}, while others include moduli~\cite{ArkaniHamed:1999dz,Dimopoulos:1996kp,Burgess:2010sy,Cicoli:2011yy}, dilatons~\cite{Damour:1994zq,Taylor:1988nw}, and DM through the Higgs portal~\cite{Piazza:2010ye}. 

In this Letter, we focus on a particular class of ultra-light scalar DM with couplings to SM fermions, $\sim \phi \, \bar{f} f$. If the SM fermion, $f$, is an electron, then this coupling induces a time variation to the electron mass, which can be searched for using atomic clocks~\cite{Arvanitaki:2015iga,Arvanitaki:2014faa,Hees:2016gop}, accelerometers~\cite{Graham:2015ifn,Stadnik:2015kia,Stadnik:2014tta}, and gravitational wave detectors~\cite{Arvanitaki:2016fyj}. In this work, we will instead explore scalar couplings between DM and SM neutrinos. These types of interactions generically result in time-varying corrections to the SM neutrino masses and/or mixing angles, which can be searched for in the unique  signals generated at neutrino oscillation experiments. Contrary to the previous studies mentioned above, the precision to which we are currently able to measure processes that are directly tied to neutrino masses does not compare to our understanding of other fundamental parameters of the SM. Regardless, as we will see below, interesting constraints can still be placed on such a scenario, given current information. We note that past studies have explored possible signals of light scalars at neutrino oscillation experiments, for example, from effects of dilatons~\cite{Halprin:1997na,Horvat:1998pp} and accelerons~\cite{Barger:2005mn,Zurek:2004vd,Kaplan:2004dq,Barger:2005mh}. Such work often relies on interactions between the light scalar and other SM fields, such as electrons and nucleons. The effects examined in this work, however, solely rely on the scalar-neutrino coupling, and to the best of our knowledge, represent the first investigation of time-varying signals at neutrino oscillation experiments from light scalar DM. 

Let us consider a model, consisting of a real scalar DM field, $\phi$, coupled to the SM neutrino mass eigenstates of the vacuum, $\nu_i$ ($i=1,2,3$), which we take to be Majorana, and whose naming convention follows that in the neutrino literature~\cite{Agashe:2014kda}. We will assume that $\phi$ couples to a pair of SM neutrinos, $\nu_1$ and $\nu_2$, in which case our effective Lagrangian (ignoring kinetic terms) takes the form
\be
\label{eq:L1}
- \mathcal{L} \supset \frac{1}{2} \, m_\phi^2 \, \phi^2 + \frac{1}{2} \, m_i \, \bar{\nu}_i \, \nu_i + g_\phi \, \phi \, \bar{\nu}_1 \, \nu_2 + \cdots
~,
\ee
where $g_\phi > 0$, an implicit sum over $i$ is assumed, and the ellipsis denotes other possible interactions with neutrinos. We remain agnostic of the origin of the vacuum masses\footnote{We have assumed that $\nu_i$ are four-component Majorana spinors, consisting of a single Weyl field, such that $m_i$ are the corresponding Majorana masses~\cite{Willenbrock:2004hu}.}, $m_i$, and assume that they are generated from some unspecified process, such as a typical seesaw mechanism~\cite{Schechter:1980gr,GellMann:1980vs,Mohapatra:1979ia,Yanagida:1979as,Minkowski:1977sc}. Diagonal couplings of the form $\phi \, \bar{\nu}_i \, \nu_i$ (as well as other off-diagonal couplings) are not forbidden, and do not spoil the interesting phenomenology considered below, but lead to less interesting signals compared to Eq.~(\ref{eq:L1}). We will comment more on this scenario towards the end of this work.\footnote{Given an ultraviolet cutoff, $\Lambda_\text{UV}$, naturalness dictates that the DM mass should satisfy $m_\phi \gtrsim g_\phi \, \Lambda_\text{UV} / 4 \pi$. In the relevant parameter space discussed below, such considerations imply that $\Lambda_\text{UV} \lesssim$ MeV. }

From a top-down perspective, $\phi$ may be identified with a dilaton of an extra-dimensional extension of the SM or a CP-violating pseudo-goldstone of a spontaneously broken global symmetry. In these theories, gauge invariance suggests that $\phi$ should also possess similar couplings to charged leptons, in which case scalar interactions with electrons often provide the best opportunity for detection. However, it is simple to construct a framework where the tree-level coupling to electrons is suppressed, for instance if $g_\phi$ is generated indirectly through a coupling between $\phi$ and a right-handed sterile neutrino, or through an appropriate choice of the ratio of vacuum expectation values in two-Higgs doublet models. However, through a loop of neutrinos, one might imagine radiatively generating a scalar coupling between $\phi$ and electrons, naturally of the form $\sim ( g_\phi \, g_2^2 / 16\, \pi^2 )  \, (m_e \, m_\nu / m_W^2)$, where $g_2$ is the $SU(2)_L$ gauge coupling. As will be shown below, for $m_\phi \sim 10^{-22} \eV$, solar neutrino detectors currently constrain DM-neutrino couplings as small as $g_\phi \sim 10^{-25}$, corresponding to radiatively generated DM-electron interactions that are several orders of magnitude below the projected reach of alternative searches~\cite{Arvanitaki:2015iga,Arvanitaki:2014faa,Hees:2016gop,Graham:2015ifn,Arvanitaki:2016fyj}. Conversely, given a detectable tree-level coupling to electrons, $\phi$ naturally has scalar interactions with neutrinos, and solar neutrino detectors serve as a complementary probe to confirm the consistency of a potential signal observed in other experiments.

%Similarly, through a loop of neutrinos, one might imagine radiatively generating a coupling between $\phi$ and electrons. As discussed above, this interaction has already been extensively studied, so for the sake of simplicity, we will assume that it is tuned to be negligibly small at low energies. 

The local number of DM particles per de Broglie wavelength cubed is $N_\phi = \rhodm / m_\phi^4 \, v^3 \sim \order{1} \, (m_\phi / 10 \eV)^{-4}$, where we have taken $v \sim 10^{-3}$ for the virialized DM velocity in the Milky Way. Since the phase space density of $\phi$ is large for $m_\phi \ll 1$ eV, it can be approximated as a non-relativistic plane wave solution to its classical equation of motion,
\be
\label{eq:EOM}
\phi (x) \simeq \frac{\sqrt{2 \, \rhodm (x)}}{m_\phi} ~ \cos{\big[ \, m_\phi  \, (t - \vec{v} \cdot \vec{x}) \, \big]} 
~,
\ee
where $|\, \vec{v}\, |$ is the virialized DM velocity, and $\rhodm (x)$ is the DM density at the spacetime coordinate $x$, depending on the situation of interest. For large enough values of $g_\phi$, the cosmological non-relativistic background of neutrinos can potentially alter the form of Eq.~(\ref{eq:EOM}). However, we find such a contribution is negligible if $\rhodm \gtrsim \big( g_\phi \, n_\nu / m_\phi \big)^2$, where $n_\nu$ is the number density of the cosmic neutrino background. At the time of recombination, this translates roughly to $g_\phi \lesssim \order{10^{-20}} \times ( m_\phi / 10^{-22} \eV )$, while for this to hold locally today we find $g_\phi \lesssim \order{10^{-13}} \times ( m_\phi / 10^{-22} \eV)$.

Neutrino oscillation experiments indirectly measure the survival probability from an initial source. For example, solar neutrino detectors infer the survival probability of electron neutrinos from the Sun, $P_{\nu_e}^\odot$, which for a fixed neutrino energy, is approximately dependent on just the solar neutrino angle, $\theta_{12}$, due to matter effects~\cite{Agashe:2014kda}. Eq.~(\ref{eq:EOM}) implies that in the presence of a coherent background $\phi$ field, the coupling $g_\phi$ of Eq.~(\ref{eq:L1}) will result in an effective shift in neutrino masses and more importantly in the mixing angles among the different weak-flavor eigenstates. To leading order in $g_\phi$, the effective form of $\theta_{12}$ picks up an additional time-oscillating term,
\be
\label{eq:th12}
\sin{\theta_{12}(t)} \simeq \sin{\theta_{12}} + \frac{\cos{\theta_{12}}}{\Delta m_{12}} ~ \frac{g_\phi \, \sqrt{2 \, \rho_{_\text{DM}}}}{m_\phi} ~ \cos{m_\phi \, t}
~,
\ee
where $\Delta m_{12} \equiv m_2 - m_1 > 0$.\footnote{Due to it's non-zero velocity, $v \sim 10^{-3}$, $\phi$ is coherent over a de Broglie wavelength and hence has a characteristic spatial variation of size $\sim 1 / ( m_\phi \, v )$. We have ignored this effect, since Eq.~(\ref{eq:EOM}) implies that relativistic neutrinos traverse a background $\phi$ field that effectively varies in time as $\sim \cos{\big[ m_\phi \, t(1 - v) \big]} \simeq  \cos{m_\phi \, t}$. }

The solar observations of Super-Kamiokande (Super-K), for example, using light water Cherenkov detectors, predominantly measure $^8B$ solar neutrinos of energy $E_\nu \sim 10 \MeV$ through electron recoil processes, $\nu + e^- \to \nu + e^-$~\cite{Fukuda:2001nj}. Since Super-K directly observes the flux of recoiled electrons, all that can be inferred is the \emph{effective} neutrino flux, defined to be
\be
\label{eq:flux}
\Phi_\text{eff} \equiv \Phi \times \left(P_{\nu_e}^\odot + \left( 1 - P_{\nu_e}^\odot \right) \frac{\sigma_{\mu, \tau}}{\sigma_e} \right)
~,
\ee
where $\Phi$ is the solar neutrino flux, and $\sigma_{\mu, \tau} / \sigma_e \sim \order{0.1}$ is the ratio of the $\nu_{\mu, \tau}-e^-$ and $\nu_e-e^-$ scattering cross sections~\cite{deGouvea:1999wg}. For energies $E_\nu > \text{few} \times \MeV$, $P_{\nu_e}^\odot \simeq \sin^2{\theta_{12}} \simeq 0.3$~\cite{Gonzalez-Garcia:2014bfa}. Therefore, barring $\order{10} \, \%$ corrections, Eqs.~(\ref{eq:th12}) and (\ref{eq:flux}) imply that $\theta_{12}(t)$ will induce an oscillating flux at solar neutrino experiments with a frequency $m_\phi$,
\be
\Phi_\text{eff} =  \Phi^{(0)} + \Phi^{(1)} \, \cos{m_\phi \, t }
~,
\ee
and an approximate  modulation fraction of
\be
\label{eq:modflux}
\frac{\Phi^{(1)}}{\Phi^{(0)}} \simeq 2 \, \cot{\theta_{12}} ~ \frac{g_\phi \, \sqrt{2 \, \rhodm}}{m_\phi \, \Delta m_{12}}
~.
\ee
For $E_\nu \lesssim 100 \keV$, the prefactor of $2 \, \cot{\theta_{12}}$ in Eq.~(\ref{eq:modflux}) should be replaced by $\sin{4 \theta_{12}}/(1 - \sin^2{2 \theta_{12}}/2)$, which is relatively suppressed by a factor of $\sim 2.5$. Hence, we will tend to focus on the more energetic $^8B$ neutrinos.
\begin{figure*}[t]
\begin{center}
\hspace{-1.8cm}\includegraphics[width=0.75\textwidth]{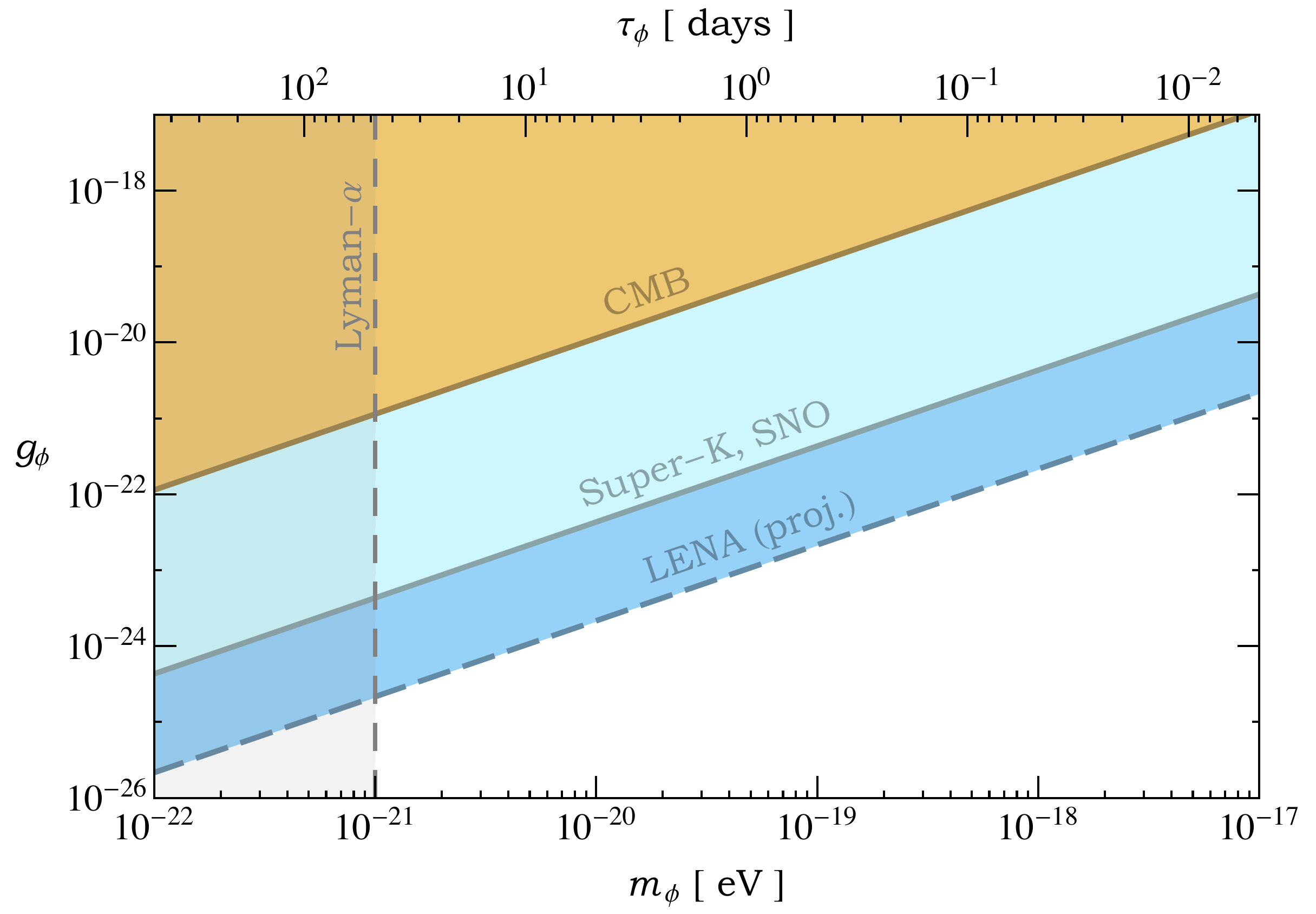} \hspace{-0.5cm}
\caption{Regions in the $m_\phi - g_\phi$ plane that are ruled out by either Planck measurements of the CMB or searches for time-varying signals in the solar neutrino data of Super-K and SNO. We also show the projected reach of the LENA liquid scintillator neutrino experiment. On the top-axis, for each value of $m_\phi$ we display the corresponding period (in days) of the modulation imprinted on the solar neutrino flux, $\tau_\phi \equiv 2 \pi / m_\phi$. In the grey shaded region, dark matter lighter than $\sim 10^{-21} \eV$ is in slight tension with observations of the Lyman-$\alpha$ forest. Current searches for anomalous periodicities in the solar neutrino data can probe DM-neutrino couplings that are orders of magnitude below limits derived from cosmological observations.}
\label{mainfig}
\end{center}
\end{figure*}

For quite some time, there have been claims of discoveries of periodicities in the solar neutrino data~\cite{Bahcall:1990dr,Sturrock:1997gp,Caldwell:2003dw,Milsztajn:2003af,Ghosh:2006me,Ghosh:2006xk,Sturrock:2003ep,Sturrock:2000jk,Sturrock:2003kv,Sturrock:2005wf,Sturrock:2004hx,Sturrock:2004jv,Sturrock:2006qz,Sturrock:2009cm,Sturrock:2012re,Sturrock:2015ivo,Desai:2016bmz}.  Different theoretical explanations have been put forward, including but not limited to various aspects of solar activity~\cite{Bahcall:1990dr,Sturrock:1997gp,Sturrock:2012re,Sturrock:2008dg} and even more exotic explanations such as a non-zero neutrino magnetic moment~\cite{Akhmedov:2002mf}. However, aside from the observed $7 \, \%$ variation due to the eccentricity of the earth's orbit~\cite{Ranucci:2007fb,Cravens:2008aa}, no such claims have been corroborated from the experimental collaborations. Super-K and the Sudbury Neutrino Observatory (SNO) found no evidence for anomalous modulations with amplitudes over $10 \, \%$ of their central flux with periods ranging from $\sim 10 \text{ minutes}-10$ years~\cite{Yoo:2003rc,Aharmim:2005iu,Collaboration:2009qz}. Furthermore, studies of proposed future technologies suggest that the liquid scintillator Low Energy Neutrino Astronomy (LENA) detector could be sensitive to variations at the $0.5 \, \%$ level~\cite{Wurm:2010mq}. Approved near-future experiments, such as SNO+, will also have increased sensitivity, although no dedicated studies of the projected reach for such signals have been performed. Eq.~(\ref{eq:modflux}) therefore implies that the null results from searches at Super-K and SNO lead to an upper bound on $g_\phi$ that approximately scales as
\be
\label{eq:gphi}
g_\phi \lesssim 4 \times 10^{-25} \times \left(\frac{m_\phi}{10^{-22} \text{ eV}}\right)
\quad \text{(solar)}
~,
\ee
where we have taken $\Delta m_{12}^2 \simeq (7.5 \pm 0.25) \times 10^{-5} \eV^2$~\cite{Gonzalez-Garcia:2014bfa} and $m_{1,2} \sim 0.1 \eV$. In the limit that $\nu_1$ is massless, this bound is weakened by approximately an order of magnitude.\footnote{Note that our estimate is conservative, since it assumes that gravitational focusing of the DM density near the Sun is negligible. Taking this into account, a simple estimate suggests that $\rhodm$ should be enhanced by a factor of $\sqrt{2 G_N M_\odot/ v^2 R_\odot} \sim \order{1}$~\cite{Alenazi:2006wu,Bromley:2011aa}. Hence, the effect on the modulation fraction is at the level of $\order{10} \, \%$.} For values of $g_\phi$ near the upper limit of Eq.~(\ref{eq:gphi}), the resulting correction to $\Delta m_{12}^2$ is at the level of $0.1\, \%$ and is a negligible effect. Such interactions are well below bounds from leptonic meson decays, which constrain couplings at the level of $g_\phi \sim 10^{-3}$~\cite{Pasquini:2015fjv}.

$\phi$ should already be present during matter-radiation equality, and hence should oscillate as in Eq.~(\ref{eq:EOM}) at the time of recombination. In the case that $\phi$ provides the dominant contribution to the effective masses of $\nu_{1,2}$, we have
\be
\label{eq:mavg}
m_{1,2} (t) \simeq g_\phi ~ \langle | \phi | \rangle \simeq \frac{2 \, g_\phi}{\pi} \frac{\sqrt{2 \, \rhodm}}{m_\phi}
~,
\ee
such that $\langle | \phi | \rangle$ is the time-averaged value of $|\phi|$. Planck measurements of the cosmic microwave background (CMB) set an upper limit on the sum of the neutrino masses at the time of photon decoupling,  $\sum_{i} m_i \lesssim 0.23 \eV$~\cite{Abazajian:2011dt,Ade:2015xua}. Eq.~(\ref{eq:mavg}) then implies that 
\be
\label{eq:CMB}
g_\phi \lesssim 10^{-22} \times \left( \frac{m_\phi}{10^{-22} \eV} \right)
\quad \text{(CMB)}
~,
\ee
which is orders of magnitude weaker than the bound from solar neutrino detectors. Projected sensitivities of CMB Stage-IV experiments could improve upon the limit in Eq.~(\ref{eq:CMB}) by approximately an order of magnitude~\cite{Wu:2014hta}.

Similar limits can be inferred from the successful prediction of Big Bang Nucleosynthesis (BBN) at temperatures $\sim 1 \MeV$~\cite{Mangano:2011ar}. During this epoch, the background density of thermal neutrinos dominates the potential of $\phi$, altering the form of Eq.~(\ref{eq:EOM}). However, since $\phi$ does not need to be present at these times, the applicability of these constraints is subject to model-dependent considerations. In light of these insights, we will assume that the temperature associated with the phase transition that sets the initial conditions for $\phi$ is significantly below $\sim 1 \MeV$. As a concrete example, if $\phi$ is produced by the misalignment mechanism, it doesn't begin to behave as cold DM until its mass overcomes the expansion rate, $m_\phi (T) \gtrsim 3 \, H(T)$, which will occur after BBN as long as $m_\phi (1 \MeV) \lesssim 10^{-15} \eV$.

In Fig.~\ref{mainfig}, we summarize our results and display the current reach in the $m_\phi-g_\phi$ plane from Planck measurements of the CMB and searches for periodicities in the solar neutrino flux from the Super-K and SNO detectors, as well as projected limits from LENA. For $10^{-22} \eV \lesssim m_\phi \lesssim 10^{-17} \eV$, the induced time-variation in the neutrino flux ranges from $\sim \order{1} \text{ year} - \, \order{10} \text{ minutes}$, respectively. Current solar neutrino experiments have focused on  modulations with periods longer than several minutes and hence do not significantly constrain larger masses, $m_\phi \gg 10^{-17} \eV$. As a result of matter effects, the probability for an electron neutrino to survive from the solar interior to the earth is approximately equal to the probability for the neutrino to survive from the solar interior to the solar surface, corresponding to a propagation time of a few seconds~\cite{Agashe:2014kda}. Therefore, in principle, DM masses as large as $m_\phi \sim 10^{-15} \eV$ can induce a detectable flux variation in solar neutrino detectors, while for $m_\phi \gg 10^{-15} \eV$ any time-dependence in Eq.~(\ref{eq:th12}) averages to a constant. In the case that $g_\phi$ is radiatively generated through the direct coupling between $\phi$ and a heavy right-handed sterile neutrino, $N$, we have the approximate relation $g_\phi \sim m_\nu / m_N$. It is then interesting to note that for $m_\phi \sim 10^{-22} \eV$ and $m_\nu \sim 0.1 \eV$, the projected reach of LENA will probe DM-neutrino couplings as small as $g_\phi \sim 10^{-26}$, corresponding to $m_N \sim 10^{16} \GeV$ and hence right-handed neutrinos at the scale of grand unified theories. Observations of the Lyman-$\alpha$ forest lead to slight tensions with dark matter masses lighter than $\sim 10^{-22} \eV$. For a recent investigation of this scenario, see, e.g., Ref.~\cite{Hui:2016ltb}. 

Compared to solar observations, atmospheric, reactor, and accelerator experiments are more directly sensitive to the neutrino mass splittings. These experiments indirectly measure survival probabilities of the form
%It is also interesting to consider other possible signals that are more directly sensitive to the neutrino mass splittings, such as observations from atmospheric, reactor, and accelerator experiments. These experiments indirectly measure survival probabilities of the form
%
\be
\label{eq:Pn}
P_\nu \simeq 1 - \sin^2{2 \theta} \, \sin^2{\frac{\Delta m^2 \, L}{4 \, E_\nu}}
~,
\ee
%. 
where $L$ is the baseline, $\Delta m^2$ is the appropriate mass-squared splitting, and $\theta$ some effective mixing angle. In this case, a simpler effective model entails coupling $\phi$ diagonally to one of the neutrino mass eigenstates, denoted as $\nu$,
\be
\label{eq:L2}
- \mathcal{L} \supset \frac{1}{2} \, g_\phi \, \phi \, \bar{\nu} \, \nu
~.
\ee
Eq.~(\ref{eq:L2}) leads to an effective time-varying mass-squared splitting between $\nu$ and any other neutrino,
\be
\label{Eq:deltam2}
\Delta m^2 (t) \simeq \langle \Delta m^2 \rangle \, + \, 2 \, g_\phi \, m_{\nu} \, \phi_0 \, \cos{m_\phi \, t}
~,
\ee
where $\phi_0 \equiv \sqrt{2 \, \rhodm} \, / \, m_\phi$, $m_\nu$ is the vacuum mass of $\nu$, and $\langle \Delta m^2 \rangle$ is the time-averaged value of the effective mass-squared splitting, $\Delta m^2 (t)$. As a result, the neutrino flux develops a time-oscillating component, of frequency $m_\phi$, and modulation fraction of
\be
\label{eq:modflux2}
\frac{\Phi^{(1)}}{\Phi^{(0)}} \simeq \frac{m_\nu \, \phi_0 \, g_\phi}{\langle P_\nu \rangle} ~ \frac{L}{2\, E_\nu}~ \sin^2{2 \theta} ~ \sin{\frac{\langle \Delta m^2 \rangle \, L}{2 \, E_\nu}} 
~,
\ee
where $\langle P_\nu \rangle$ is defined as in Eq.~(\ref{eq:Pn}), but with $\Delta m^2$ replaced by $\langle \Delta m^2 \rangle$. 

As an example of the applicability of Eq.~(\ref{eq:modflux2}), let us consider KamLAND, a liquid scintillator detector, which is designed to study the anti-neutrino flux of nuclear power reactors though the process $\bar{\nu}_e + p \to e^+ + n$ and provides an independent determination of $\theta_{12}$ and $\Delta m_{12}^2$. We estimate that DM-neutrino couplings at the level of $g_\phi \sim 10^{-24} \times (m_\phi / 10^{-22} \eV)$ would generate an $\order{10} \, \%$ modulation fraction in the fluxes as measured by KamLAND. It has been suggested that this level of time-dependence could be detectable~\cite{Fogli:2005qa,Lisi:2004jw}, however, such an analysis would involve disentangling this effect from the large time variation in the nuclear reactor fission rates and certainly would provide weaker limits compared to solar neutrino observations. 

In this Letter, we have considered a class of models involving small couplings between an ultra-light scalar DM field and SM neutrinos. These types of interactions generically modify the neutrino mass splittings and mixing angles to include additional components that vary in time periodically with a frequency and amplitude dictated by the mass and energy density of the DM. Solar neutrino detectors such as Super-K, SNO, and LENA constitute a particularly interesting class of experiments in regards to these types of  models. Null results from recent searches for anomalous periodicities in the solar neutrino flux strongly constrain the DM-neutrino coupling to be orders of magnitude below current and projected limits derived from observations of the effective number of relativistic species during recombination. Although we have focused on solar neutrino experiments, other types of detectors that measure the flux from atmospheric, reactor, and accelerator sources may also have sensitivity to similar phenomenology.

\section*{Acknowledgments}
We would like to thank Dan Hooper, Andrew Long, Gordan Krnjaic, Peter Graham, and Andr\'e de Gouv\^ea for useful conversations. AB is supported by the Kavli Institute for Cosmological Physics at the University of Chicago through grant NSF PHY-1125897.

\bibliography{neutrino}

%Merlin.mbs v4.21 2009-07-09.
\begin{thebibliography}{10}%
\makeatletter
\providecommand \@ifxundefined [1]{%
 \ifx #1\undefined \expandafter \@firstoftwo
 \else \expandafter \@secondoftwo
\fi
}%
\providecommand \@ifnum [1]{%
 \ifnum #1\expandafter \@firstoftwo
 \else \expandafter \@secondoftwo
\fi
}%
\providecommand \enquote [1]{``#1''}%
\providecommand \bibnamefont  [1]{#1}%
\providecommand \bibfnamefont [1]{#1}%
\providecommand \citenamefont [1]{#1}%
\providecommand\href[0]{\@sanitize\@href}%
\providecommand\@href[1]{\endgroup\@@startlink{#1}\endgroup\@@href}%
\providecommand\@@href[1]{#1\@@endlink}%
\providecommand \@sanitize [0]{\begingroup\catcode`\&12\catcode`\#12\relax}%
\@ifxundefined \pdfoutput {\@firstoftwo}{%
 \@ifnum{\z@=\pdfoutput}{\@firstoftwo}{\@secondoftwo}%
}{%
 \providecommand\@@startlink[1]{\leavevmode\special{html:<a href="#1">}}%
 \providecommand\@@endlink[0]{\special{html:</a>}}%
}{%
 \providecommand\@@startlink[1]{%
  \leavevmode
  \pdfstartlink
   attr{/Border[0 0 1 ]/H/I/C[0 1 1]}%
   user{/Subtype/Link/A<</Type/Action/S/URI/URI(#1)>>}%
  \relax
 }%
 \providecommand\@@endlink[0]{\pdfendlink}%
}%
\providecommand \url  [0]{\begingroup\@sanitize \@url }%
\providecommand \@url [1]{\endgroup\@href {#1}{\urlprefix}}%
\providecommand \urlprefix [0]{URL }%
\providecommand \Eprint[0]{\href }%
\@ifxundefined \urlstyle {%
  \providecommand \doi [1]{doi:\discretionary{}{}{}#1}%
}{%
  \providecommand \doi [0]{doi:\discretionary{}{}{}\begingroup
  \urlstyle{rm}\Url }%
}%
\providecommand \doibase [0]{http://dx.doi.org/}%
\providecommand \Doi[1]{\href{\doibase#1}}%
\providecommand \bibAnnote [3]{%
  \BibitemShut{#1}%
  \begin{quotation}\noindent
    \textsc{Key:}\ #2\\\textsc{Annotation:}\ #3%
  \end{quotation}%
}%
\providecommand \bibAnnoteFile [2]{%
  \IfFileExists{#2}{\bibAnnote {#1} {#2} {\input{#2}}}{}%
}%
\providecommand \typeout [0]{\immediate \write \m@ne }%
\providecommand \selectlanguage [0]{\@gobble}%
\providecommand \bibinfo [0]{\@secondoftwo}%
\providecommand \bibfield [0]{\@secondoftwo}%
\providecommand \translation [1]{[#1]}%
\providecommand \BibitemOpen[0]{}%
\providecommand \bibitemStop [0]{}%
\providecommand \bibitemNoStop [0]{.\EOS\space}%
\providecommand \EOS [0]{\spacefactor3000\relax}%
\providecommand \BibitemShut [1]{\csname bibitem#1\endcsname}%
%</preamble>
\bibitem{Akerib:2015rjg}%
  \BibitemOpen
  \bibfield{author}{%
  \bibinfo {author} {\bibfnamefont{D.~S.}\ \bibnamefont{Akerib}} \emph{et~al.}
  (\bibinfo {collaboration} {LUX}),\ }%
  \bibfield{journal}{%
  \Doi{10.1103/PhysRevLett.116.161301}{\bibinfo {journal} {Phys. Rev. Lett.}}\
  }%
  \textbf{\bibinfo {volume} {116}},\ \bibinfo {pages} {161301} (\bibinfo {year}
  {2016}),\ \Eprint{http://arxiv.org/abs/1512.03506}{arXiv:1512.03506
  [astro-ph.CO]}%
  \bibAnnoteFile{NoStop}{Akerib:2015rjg}%
%%CITATION = ARXIV:1512.03506;%%
\bibitem{Aprile:2012nq}%
  \BibitemOpen
  \bibfield{author}{%
  \bibinfo {author} {\bibfnamefont{E.}~\bibnamefont{Aprile}} \emph{et~al.}
  (\bibinfo {collaboration} {XENON100}),\ }%
  \bibfield{journal}{%
  \Doi{10.1103/PhysRevLett.109.181301}{\bibinfo {journal} {Phys. Rev. Lett.}}\
  }%
  \textbf{\bibinfo {volume} {109}},\ \bibinfo {pages} {181301} (\bibinfo {year}
  {2012}),\ \Eprint{http://arxiv.org/abs/1207.5988}{arXiv:1207.5988
  [astro-ph.CO]}%
  \bibAnnoteFile{NoStop}{Aprile:2012nq}%
%%CITATION = ARXIV:1207.5988;%%
\bibitem{Tan:2016zwf}%
  \BibitemOpen
  \bibfield{author}{%
  \bibinfo {author} {\bibfnamefont{A.}~\bibnamefont{Tan}} \emph{et~al.}}%
   (\bibinfo {year} {2016}),\
  \Eprint{http://arxiv.org/abs/1607.07400}{arXiv:1607.07400 [hep-ex]}%
  \bibAnnoteFile{NoStop}{Tan:2016zwf}%
%%CITATION = ARXIV:1607.07400;%%
\bibitem{Agnese:2013jaa}%
  \BibitemOpen
  \bibfield{author}{%
  \bibinfo {author} {\bibfnamefont{R.}~\bibnamefont{Agnese}} \emph{et~al.}
  (\bibinfo {collaboration} {SuperCDMS}),\ }%
  \bibfield{journal}{%
  \Doi{10.1103/PhysRevLett.112.041302}{\bibinfo {journal} {Phys. Rev. Lett.}}\
  }%
  \textbf{\bibinfo {volume} {112}},\ \bibinfo {pages} {041302} (\bibinfo {year}
  {2014}),\ \Eprint{http://arxiv.org/abs/1309.3259}{arXiv:1309.3259
  [physics.ins-det]}%
  \bibAnnoteFile{NoStop}{Agnese:2013jaa}%
%%CITATION = ARXIV:1309.3259;%%
\bibitem{Aprile:2015uzo}%
  \BibitemOpen
  \bibfield{author}{%
  \bibinfo {author} {\bibfnamefont{E.}~\bibnamefont{Aprile}} \emph{et~al.}
  (\bibinfo {collaboration} {XENON}),\ }%
  \bibfield{journal}{%
  \Doi{10.1088/1475-7516/2016/04/027}{\bibinfo {journal} {JCAP}}\ }%
  \textbf{\bibinfo {volume} {1604}},\ \bibinfo {pages} {027} (\bibinfo {year}
  {2016}),\ \Eprint{http://arxiv.org/abs/1512.07501}{arXiv:1512.07501
  [physics.ins-det]}%
  \bibAnnoteFile{NoStop}{Aprile:2015uzo}%
%%CITATION = ARXIV:1512.07501;%%
\bibitem{Akerib:2015cja}%
  \BibitemOpen
  \bibfield{author}{%
  \bibinfo {author} {\bibfnamefont{D.~S.}\ \bibnamefont{Akerib}} \emph{et~al.}
  (\bibinfo {collaboration} {LZ})}%
   (\bibinfo {year} {2015}),\
  \Eprint{http://arxiv.org/abs/1509.02910}{arXiv:1509.02910 [physics.ins-det]}%
  \bibAnnoteFile{NoStop}{Akerib:2015cja}%
%%CITATION = ARXIV:1509.02910;%%
\bibitem{Izaguirre:2015yja}%
  \BibitemOpen
  \bibfield{author}{%
  \bibinfo {author} {\bibfnamefont{E.}~\bibnamefont{Izaguirre}}, \bibinfo
  {author} {\bibfnamefont{G.}~\bibnamefont{Krnjaic}}, \bibinfo {author}
  {\bibfnamefont{P.}~\bibnamefont{Schuster}},\ and\ \bibinfo {author}
  {\bibfnamefont{N.}~\bibnamefont{Toro}},\ }%
  \bibfield{journal}{%
  \Doi{10.1103/PhysRevLett.115.251301}{\bibinfo {journal} {Phys. Rev. Lett.}}\
  }%
  \textbf{\bibinfo {volume} {115}},\ \bibinfo {pages} {251301} (\bibinfo {year}
  {2015}),\ \Eprint{http://arxiv.org/abs/1505.00011}{arXiv:1505.00011
  [hep-ph]}%
  \bibAnnoteFile{NoStop}{Izaguirre:2015yja}%
%%CITATION = ARXIV:1505.00011;%%
\bibitem{Kaplan:2009ag}%
  \BibitemOpen
  \bibfield{author}{%
  \bibinfo {author} {\bibfnamefont{D.~E.}\ \bibnamefont{Kaplan}}, \bibinfo
  {author} {\bibfnamefont{M.~A.}\ \bibnamefont{Luty}},\ and\ \bibinfo {author}
  {\bibfnamefont{K.~M.}\ \bibnamefont{Zurek}},\ }%
  \bibfield{journal}{%
  \Doi{10.1103/PhysRevD.79.115016}{\bibinfo {journal} {Phys. Rev.}}\ }%
  \textbf{\bibinfo {volume} {D79}},\ \bibinfo {pages} {115016} (\bibinfo {year}
  {2009}),\ \Eprint{http://arxiv.org/abs/0901.4117}{arXiv:0901.4117 [hep-ph]}%
  \bibAnnoteFile{NoStop}{Kaplan:2009ag}%
%%CITATION = ARXIV:0901.4117;%%
\bibitem{Boehm:2003hm}%
  \BibitemOpen
  \bibfield{author}{%
  \bibinfo {author} {\bibfnamefont{C.}~\bibnamefont{Boehm}}\ and\ \bibinfo
  {author} {\bibfnamefont{P.}~\bibnamefont{Fayet}},\ }%
  \bibfield{journal}{%
  \Doi{10.1016/j.nuclphysb.2004.01.015}{\bibinfo {journal} {Nucl. Phys.}}\ }%
  \textbf{\bibinfo {volume} {B683}},\ \bibinfo {pages} {219} (\bibinfo {year}
  {2004}),\ \Eprint{http://arxiv.org/abs/hep-ph/0305261}{arXiv:hep-ph/0305261
  [hep-ph]}%
  \bibAnnoteFile{NoStop}{Boehm:2003hm}%
%%CITATION = HEP-PH/0305261;%%
\bibitem{Pospelov:2007mp}%
  \BibitemOpen
  \bibfield{author}{%
  \bibinfo {author} {\bibfnamefont{M.}~\bibnamefont{Pospelov}}, \bibinfo
  {author} {\bibfnamefont{A.}~\bibnamefont{Ritz}},\ and\ \bibinfo {author}
  {\bibfnamefont{M.~B.}\ \bibnamefont{Voloshin}},\ }%
  \bibfield{journal}{%
  \Doi{10.1016/j.physletb.2008.02.052}{\bibinfo {journal} {Phys. Lett.}}\ }%
  \textbf{\bibinfo {volume} {B662}},\ \bibinfo {pages} {53} (\bibinfo {year}
  {2008}),\ \Eprint{http://arxiv.org/abs/0711.4866}{arXiv:0711.4866 [hep-ph]}%
  \bibAnnoteFile{NoStop}{Pospelov:2007mp}%
%%CITATION = ARXIV:0711.4866;%%
\bibitem{Hooper:2008im}%
  \BibitemOpen
  \bibfield{author}{%
  \bibinfo {author} {\bibfnamefont{D.}~\bibnamefont{Hooper}}\ and\ \bibinfo
  {author} {\bibfnamefont{K.~M.}\ \bibnamefont{Zurek}},\ }%
  \bibfield{journal}{%
  \Doi{10.1103/PhysRevD.77.087302}{\bibinfo {journal} {Phys. Rev.}}\ }%
  \textbf{\bibinfo {volume} {D77}},\ \bibinfo {pages} {087302} (\bibinfo {year}
  {2008}),\ \Eprint{http://arxiv.org/abs/0801.3686}{arXiv:0801.3686 [hep-ph]}%
  \bibAnnoteFile{NoStop}{Hooper:2008im}%
%%CITATION = ARXIV:0801.3686;%%
\bibitem{Feng:2008ya}%
  \BibitemOpen
  \bibfield{author}{%
  \bibinfo {author} {\bibfnamefont{J.~L.}\ \bibnamefont{Feng}}\ and\ \bibinfo
  {author} {\bibfnamefont{J.}~\bibnamefont{Kumar}},\ }%
  \bibfield{journal}{%
  \Doi{10.1103/PhysRevLett.101.231301}{\bibinfo {journal} {Phys. Rev. Lett.}}\
  }%
  \textbf{\bibinfo {volume} {101}},\ \bibinfo {pages} {231301} (\bibinfo {year}
  {2008}),\ \Eprint{http://arxiv.org/abs/0803.4196}{arXiv:0803.4196 [hep-ph]}%
  \bibAnnoteFile{NoStop}{Feng:2008ya}%
%%CITATION = ARXIV:0803.4196;%%
\bibitem{Morrissey:2009ur}%
  \BibitemOpen
  \bibfield{author}{%
  \bibinfo {author} {\bibfnamefont{D.~E.}\ \bibnamefont{Morrissey}}, \bibinfo
  {author} {\bibfnamefont{D.}~\bibnamefont{Poland}},\ and\ \bibinfo {author}
  {\bibfnamefont{K.~M.}\ \bibnamefont{Zurek}},\ }%
  \bibfield{journal}{%
  \Doi{10.1088/1126-6708/2009/07/050}{\bibinfo {journal} {JHEP}}\ }%
  \textbf{\bibinfo {volume} {07}},\ \bibinfo {pages} {050} (\bibinfo {year}
  {2009}),\ \Eprint{http://arxiv.org/abs/0904.2567}{arXiv:0904.2567 [hep-ph]}%
  \bibAnnoteFile{NoStop}{Morrissey:2009ur}%
%%CITATION = ARXIV:0904.2567;%%
\bibitem{Hochberg:2014dra}%
  \BibitemOpen
  \bibfield{author}{%
  \bibinfo {author} {\bibfnamefont{Y.}~\bibnamefont{Hochberg}}, \bibinfo
  {author} {\bibfnamefont{E.}~\bibnamefont{Kuflik}}, \bibinfo {author}
  {\bibfnamefont{T.}~\bibnamefont{Volansky}},\ and\ \bibinfo {author}
  {\bibfnamefont{J.~G.}\ \bibnamefont{Wacker}},\ }%
  \bibfield{journal}{%
  \Doi{10.1103/PhysRevLett.113.171301}{\bibinfo {journal} {Phys. Rev. Lett.}}\
  }%
  \textbf{\bibinfo {volume} {113}},\ \bibinfo {pages} {171301} (\bibinfo {year}
  {2014}),\ \Eprint{http://arxiv.org/abs/1402.5143}{arXiv:1402.5143 [hep-ph]}%
  \bibAnnoteFile{NoStop}{Hochberg:2014dra}%
%%CITATION = ARXIV:1402.5143;%%
\bibitem{Hochberg:2014kqa}%
  \BibitemOpen
  \bibfield{author}{%
  \bibinfo {author} {\bibfnamefont{Y.}~\bibnamefont{Hochberg}}, \bibinfo
  {author} {\bibfnamefont{E.}~\bibnamefont{Kuflik}}, \bibinfo {author}
  {\bibfnamefont{H.}~\bibnamefont{Murayama}}, \bibinfo {author}
  {\bibfnamefont{T.}~\bibnamefont{Volansky}},\ and\ \bibinfo {author}
  {\bibfnamefont{J.~G.}\ \bibnamefont{Wacker}},\ }%
  \bibfield{journal}{%
  \Doi{10.1103/PhysRevLett.115.021301}{\bibinfo {journal} {Phys. Rev. Lett.}}\
  }%
  \textbf{\bibinfo {volume} {115}},\ \bibinfo {pages} {021301} (\bibinfo {year}
  {2015}),\ \Eprint{http://arxiv.org/abs/1411.3727}{arXiv:1411.3727 [hep-ph]}%
  \bibAnnoteFile{NoStop}{Hochberg:2014kqa}%
%%CITATION = ARXIV:1411.3727;%%
\bibitem{Schutz:2016tid}%
  \BibitemOpen
  \bibfield{author}{%
  \bibinfo {author} {\bibfnamefont{K.}~\bibnamefont{Schutz}}\ and\ \bibinfo
  {author} {\bibfnamefont{K.~M.}\ \bibnamefont{Zurek}}}%
   (\bibinfo {year} {2016}),\
  \Eprint{http://arxiv.org/abs/1604.08206}{arXiv:1604.08206 [hep-ph]}%
  \bibAnnoteFile{NoStop}{Schutz:2016tid}%
%%CITATION = ARXIV:1604.08206;%%
\bibitem{Hochberg:2015fth}%
  \BibitemOpen
  \bibfield{author}{%
  \bibinfo {author} {\bibfnamefont{Y.}~\bibnamefont{Hochberg}}, \bibinfo
  {author} {\bibfnamefont{M.}~\bibnamefont{Pyle}}, \bibinfo {author}
  {\bibfnamefont{Y.}~\bibnamefont{Zhao}},\ and\ \bibinfo {author}
  {\bibfnamefont{K.~M.}\ \bibnamefont{Zurek}}}%
   (\bibinfo {year} {2015}),\
  \Eprint{http://arxiv.org/abs/1512.04533}{arXiv:1512.04533 [hep-ph]}%
  \bibAnnoteFile{NoStop}{Hochberg:2015fth}%
%%CITATION = ARXIV:1512.04533;%%
\bibitem{Hochberg:2015pha}%
  \BibitemOpen
  \bibfield{author}{%
  \bibinfo {author} {\bibfnamefont{Y.}~\bibnamefont{Hochberg}}, \bibinfo
  {author} {\bibfnamefont{Y.}~\bibnamefont{Zhao}},\ and\ \bibinfo {author}
  {\bibfnamefont{K.~M.}\ \bibnamefont{Zurek}},\ }%
  \bibfield{journal}{%
  \Doi{10.1103/PhysRevLett.116.011301}{\bibinfo {journal} {Phys. Rev. Lett.}}\
  }%
  \textbf{\bibinfo {volume} {116}},\ \bibinfo {pages} {011301} (\bibinfo {year}
  {2016}),\ \Eprint{http://arxiv.org/abs/1504.07237}{arXiv:1504.07237
  [hep-ph]}%
  \bibAnnoteFile{NoStop}{Hochberg:2015pha}%
%%CITATION = ARXIV:1504.07237;%%
\bibitem{Khlopov:1985jw}%
  \BibitemOpen
  \bibfield{author}{%
  \bibinfo {author} {\bibfnamefont{M.}~\bibnamefont{Khlopov}}, \bibinfo
  {author} {\bibfnamefont{B.~A.}\ \bibnamefont{Malomed}},\ and\ \bibinfo
  {author} {\bibfnamefont{I.~B.}\ \bibnamefont{Zeldovich}},\ }%
  \bibfield{journal}{%
  \bibinfo {journal} {Mon. Not. Roy. Astron. Soc.}\ }%
  \textbf{\bibinfo {volume} {215}},\ \bibinfo {pages} {575} (\bibinfo {year}
  {1985})%
  \bibAnnoteFile{NoStop}{Khlopov:1985jw}%
%%CITATION = MNRAA,215,575;%%
\bibitem{Press:1989id}%
  \BibitemOpen
  \bibfield{author}{%
  \bibinfo {author} {\bibfnamefont{W.~H.}\ \bibnamefont{Press}}, \bibinfo
  {author} {\bibfnamefont{B.~S.}\ \bibnamefont{Ryden}},\ and\ \bibinfo {author}
  {\bibfnamefont{D.~N.}\ \bibnamefont{Spergel}},\ }%
  \bibfield{journal}{%
  \Doi{10.1103/PhysRevLett.64.1084}{\bibinfo {journal} {Phys. Rev. Lett.}}\ }%
  \textbf{\bibinfo {volume} {64}},\ \bibinfo {pages} {1084} (\bibinfo {year}
  {1990})%
  \bibAnnoteFile{NoStop}{Press:1989id}%
%%CITATION = PRLTA,64,1084;%%
\bibitem{Sahni:1999qe}%
  \BibitemOpen
  \bibfield{author}{%
  \bibinfo {author} {\bibfnamefont{V.}~\bibnamefont{Sahni}}\ and\ \bibinfo
  {author} {\bibfnamefont{L.-M.}\ \bibnamefont{Wang}},\ }%
  \bibfield{journal}{%
  \Doi{10.1103/PhysRevD.62.103517}{\bibinfo {journal} {Phys. Rev.}}\ }%
  \textbf{\bibinfo {volume} {D62}},\ \bibinfo {pages} {103517} (\bibinfo {year}
  {2000}),\
  \Eprint{http://arxiv.org/abs/astro-ph/9910097}{arXiv:astro-ph/9910097
  [astro-ph]}%
  \bibAnnoteFile{NoStop}{Sahni:1999qe}%
%%CITATION = ASTRO-PH/9910097;%%
\bibitem{Hu:2000ke}%
  \BibitemOpen
  \bibfield{author}{%
  \bibinfo {author} {\bibfnamefont{W.}~\bibnamefont{Hu}}, \bibinfo {author}
  {\bibfnamefont{R.}~\bibnamefont{Barkana}},\ and\ \bibinfo {author}
  {\bibfnamefont{A.}~\bibnamefont{Gruzinov}},\ }%
  \bibfield{journal}{%
  \Doi{10.1103/PhysRevLett.85.1158}{\bibinfo {journal} {Phys. Rev. Lett.}}\ }%
  \textbf{\bibinfo {volume} {85}},\ \bibinfo {pages} {1158} (\bibinfo {year}
  {2000}),\
  \Eprint{http://arxiv.org/abs/astro-ph/0003365}{arXiv:astro-ph/0003365
  [astro-ph]}%
  \bibAnnoteFile{NoStop}{Hu:2000ke}%
%%CITATION = ASTRO-PH/0003365;%%
\bibitem{Marsh:2015xka}%
  \BibitemOpen
  \bibfield{author}{%
  \bibinfo {author} {\bibfnamefont{D.~J.~E.}\ \bibnamefont{Marsh}},\ }%
  \bibfield{journal}{%
  \Doi{10.1016/j.physrep.2016.06.005}{\bibinfo {journal} {Phys. Rept.}}\ }%
  \textbf{\bibinfo {volume} {643}},\ \bibinfo {pages} {1} (\bibinfo {year}
  {2016}),\ \Eprint{http://arxiv.org/abs/1510.07633}{arXiv:1510.07633
  [astro-ph.CO]}%
  \bibAnnoteFile{NoStop}{Marsh:2015xka}%
%%CITATION = ARXIV:1510.07633;%%
\bibitem{Hlozek:2014lca}%
  \BibitemOpen
  \bibfield{author}{%
  \bibinfo {author} {\bibfnamefont{R.}~\bibnamefont{Hlozek}}, \bibinfo {author}
  {\bibfnamefont{D.}~\bibnamefont{Grin}}, \bibinfo {author}
  {\bibfnamefont{D.~J.~E.}\ \bibnamefont{Marsh}},\ and\ \bibinfo {author}
  {\bibfnamefont{P.~G.}\ \bibnamefont{Ferreira}},\ }%
  \bibfield{journal}{%
  \Doi{10.1103/PhysRevD.91.103512}{\bibinfo {journal} {Phys. Rev.}}\ }%
  \textbf{\bibinfo {volume} {D91}},\ \bibinfo {pages} {103512} (\bibinfo {year}
  {2015}),\ \Eprint{http://arxiv.org/abs/1410.2896}{arXiv:1410.2896
  [astro-ph.CO]}%
  \bibAnnoteFile{NoStop}{Hlozek:2014lca}%
%%CITATION = ARXIV:1410.2896;%%
\bibitem{Bozek:2014uqa}%
  \BibitemOpen
  \bibfield{author}{%
  \bibinfo {author} {\bibfnamefont{B.}~\bibnamefont{Bozek}}, \bibinfo {author}
  {\bibfnamefont{D.~J.~E.}\ \bibnamefont{Marsh}}, \bibinfo {author}
  {\bibfnamefont{J.}~\bibnamefont{Silk}},\ and\ \bibinfo {author}
  {\bibfnamefont{R.~F.~G.}\ \bibnamefont{Wyse}},\ }%
  \bibfield{journal}{%
  \Doi{10.1093/mnras/stv624}{\bibinfo {journal} {Mon. Not. Roy. Astron. Soc.}}\
  }%
  \textbf{\bibinfo {volume} {450}},\ \bibinfo {pages} {209} (\bibinfo {year}
  {2015}),\ \Eprint{http://arxiv.org/abs/1409.3544}{arXiv:1409.3544
  [astro-ph.CO]}%
  \bibAnnoteFile{NoStop}{Bozek:2014uqa}%
%%CITATION = ARXIV:1409.3544;%%
\bibitem{Sarkar:2015dib}%
  \BibitemOpen
  \bibfield{author}{%
  \bibinfo {author} {\bibfnamefont{A.}~\bibnamefont{Sarkar}}, \bibinfo {author}
  {\bibfnamefont{R.}~\bibnamefont{Mondal}}, \bibinfo {author}
  {\bibfnamefont{S.}~\bibnamefont{Das}}, \bibinfo {author}
  {\bibfnamefont{S.}~\bibnamefont{Sethi}}, \bibinfo {author}
  {\bibfnamefont{S.}~\bibnamefont{Bharadwaj}},\ and\ \bibinfo {author}
  {\bibfnamefont{D.~J.~E.}\ \bibnamefont{Marsh}},\ }%
  \bibfield{journal}{%
  \Doi{10.1088/1475-7516/2016/04/012}{\bibinfo {journal} {JCAP}}\ }%
  \textbf{\bibinfo {volume} {1604}},\ \bibinfo {pages} {012} (\bibinfo {year}
  {2016}),\ \Eprint{http://arxiv.org/abs/1512.03325}{arXiv:1512.03325
  [astro-ph.CO]}%
  \bibAnnoteFile{NoStop}{Sarkar:2015dib}%
%%CITATION = ARXIV:1512.03325;%%
\bibitem{Schive:2014dra}%
  \BibitemOpen
  \bibfield{author}{%
  \bibinfo {author} {\bibfnamefont{H.-Y.}\ \bibnamefont{Schive}}, \bibinfo
  {author} {\bibfnamefont{T.}~\bibnamefont{Chiueh}},\ and\ \bibinfo {author}
  {\bibfnamefont{T.}~\bibnamefont{Broadhurst}},\ }%
  \bibfield{journal}{%
  \Doi{10.1038/nphys2996}{\bibinfo {journal} {Nature Phys.}}\ }%
  \textbf{\bibinfo {volume} {10}},\ \bibinfo {pages} {496} (\bibinfo {year}
  {2014}),\ \Eprint{http://arxiv.org/abs/1406.6586}{arXiv:1406.6586
  [astro-ph.GA]}%
  \bibAnnoteFile{NoStop}{Schive:2014dra}%
%%CITATION = ARXIV:1406.6586;%%
\bibitem{Hlozek:2016lzm}%
  \BibitemOpen
  \bibfield{author}{%
  \bibinfo {author} {\bibfnamefont{R.}~\bibnamefont{Hlo?ek}}, \bibinfo {author}
  {\bibfnamefont{D.~J.~E.}\ \bibnamefont{Marsh}}, \bibinfo {author}
  {\bibfnamefont{D.}~\bibnamefont{Grin}}, \bibinfo {author}
  {\bibfnamefont{R.}~\bibnamefont{Allison}}, \bibinfo {author}
  {\bibfnamefont{J.}~\bibnamefont{Dunkley}},\ and\ \bibinfo {author}
  {\bibfnamefont{E.}~\bibnamefont{Calabrese}}}%
   (\bibinfo {year} {2016}),\
  \Eprint{http://arxiv.org/abs/1607.08208}{arXiv:1607.08208 [astro-ph.CO]}%
  \bibAnnoteFile{NoStop}{Hlozek:2016lzm}%
%%CITATION = ARXIV:1607.08208;%%
\bibitem{Amendola:2005ad}%
  \BibitemOpen
  \bibfield{author}{%
  \bibinfo {author} {\bibfnamefont{L.}~\bibnamefont{Amendola}}\ and\ \bibinfo
  {author} {\bibfnamefont{R.}~\bibnamefont{Barbieri}},\ }%
  \bibfield{journal}{%
  \Doi{10.1016/j.physletb.2006.08.069}{\bibinfo {journal} {Phys. Lett.}}\ }%
  \textbf{\bibinfo {volume} {B642}},\ \bibinfo {pages} {192} (\bibinfo {year}
  {2006}),\ \Eprint{http://arxiv.org/abs/hep-ph/0509257}{arXiv:hep-ph/0509257
  [hep-ph]}%
  \bibAnnoteFile{NoStop}{Amendola:2005ad}%
%%CITATION = HEP-PH/0509257;%%
\bibitem{Peccei:1977hh}%
  \BibitemOpen
  \bibfield{author}{%
  \bibinfo {author} {\bibfnamefont{R.~D.}\ \bibnamefont{Peccei}}\ and\ \bibinfo
  {author} {\bibfnamefont{H.~R.}\ \bibnamefont{Quinn}},\ }%
  \bibfield{journal}{%
  \Doi{10.1103/PhysRevLett.38.1440}{\bibinfo {journal} {Phys. Rev. Lett.}}\ }%
  \textbf{\bibinfo {volume} {38}},\ \bibinfo {pages} {1440} (\bibinfo {year}
  {1977})%
  \bibAnnoteFile{NoStop}{Peccei:1977hh}%
%%CITATION = PRLTA,38,1440;%%
\bibitem{Weinberg:1977ma}%
  \BibitemOpen
  \bibfield{author}{%
  \bibinfo {author} {\bibfnamefont{S.}~\bibnamefont{Weinberg}},\ }%
  \bibfield{journal}{%
  \Doi{10.1103/PhysRevLett.40.223}{\bibinfo {journal} {Phys. Rev. Lett.}}\ }%
  \textbf{\bibinfo {volume} {40}},\ \bibinfo {pages} {223} (\bibinfo {year}
  {1978})%
  \bibAnnoteFile{NoStop}{Weinberg:1977ma}%
%%CITATION = PRLTA,40,223;%%
\bibitem{Wilczek:1977pj}%
  \BibitemOpen
  \bibfield{author}{%
  \bibinfo {author} {\bibfnamefont{F.}~\bibnamefont{Wilczek}},\ }%
  \bibfield{journal}{%
  \Doi{10.1103/PhysRevLett.40.279}{\bibinfo {journal} {Phys. Rev. Lett.}}\ }%
  \textbf{\bibinfo {volume} {40}},\ \bibinfo {pages} {279} (\bibinfo {year}
  {1978})%
  \bibAnnoteFile{NoStop}{Wilczek:1977pj}%
%%CITATION = PRLTA,40,279;%%
\bibitem{ArkaniHamed:1999dz}%
  \BibitemOpen
  \bibfield{author}{%
  \bibinfo {author} {\bibfnamefont{N.}~\bibnamefont{Arkani-Hamed}}, \bibinfo
  {author} {\bibfnamefont{L.~J.}\ \bibnamefont{Hall}}, \bibinfo {author}
  {\bibfnamefont{D.}~\bibnamefont{Tucker-Smith}},\ and\ \bibinfo {author}
  {\bibfnamefont{N.}~\bibnamefont{Weiner}},\ }%
  \bibfield{journal}{%
  \Doi{10.1103/PhysRevD.62.105002}{\bibinfo {journal} {Phys. Rev.}}\ }%
  \textbf{\bibinfo {volume} {D62}},\ \bibinfo {pages} {105002} (\bibinfo {year}
  {2000}),\ \Eprint{http://arxiv.org/abs/hep-ph/9912453}{arXiv:hep-ph/9912453
  [hep-ph]}%
  \bibAnnoteFile{NoStop}{ArkaniHamed:1999dz}%
%%CITATION = HEP-PH/9912453;%%
\bibitem{Dimopoulos:1996kp}%
  \BibitemOpen
  \bibfield{author}{%
  \bibinfo {author} {\bibfnamefont{S.}~\bibnamefont{Dimopoulos}}\ and\ \bibinfo
  {author} {\bibfnamefont{G.~F.}\ \bibnamefont{Giudice}},\ }%
  \bibfield{booktitle}{%
  \emph{\bibinfo {booktitle} {{ITP Workshop on SUSY Phenomena and SUSY GUTS
  Santa Barbara, California, December 7-9, 1995}}},\ }%
  \bibfield{journal}{%
  \Doi{10.1016/0370-2693(96)00390-5}{\bibinfo {journal} {Phys. Lett.}}\ }%
  \textbf{\bibinfo {volume} {B379}},\ \bibinfo {pages} {105} (\bibinfo {year}
  {1996}),\ \Eprint{http://arxiv.org/abs/hep-ph/9602350}{arXiv:hep-ph/9602350
  [hep-ph]}%
  \bibAnnoteFile{NoStop}{Dimopoulos:1996kp}%
%%CITATION = HEP-PH/9602350;%%
\bibitem{Burgess:2010sy}%
  \BibitemOpen
  \bibfield{author}{%
  \bibinfo {author} {\bibfnamefont{C.~P.}\ \bibnamefont{Burgess}}, \bibinfo
  {author} {\bibfnamefont{A.}~\bibnamefont{Maharana}},\ and\ \bibinfo {author}
  {\bibfnamefont{F.}~\bibnamefont{Quevedo}},\ }%
  \bibfield{journal}{%
  \Doi{10.1007/JHEP05(2011)010}{\bibinfo {journal} {JHEP}}\ }%
  \textbf{\bibinfo {volume} {05}},\ \bibinfo {pages} {010} (\bibinfo {year}
  {2011}),\ \Eprint{http://arxiv.org/abs/1005.1199}{arXiv:1005.1199 [hep-th]}%
  \bibAnnoteFile{NoStop}{Burgess:2010sy}%
%%CITATION = ARXIV:1005.1199;%%
\bibitem{Cicoli:2011yy}%
  \BibitemOpen
  \bibfield{author}{%
  \bibinfo {author} {\bibfnamefont{M.}~\bibnamefont{Cicoli}}, \bibinfo {author}
  {\bibfnamefont{C.~P.}\ \bibnamefont{Burgess}},\ and\ \bibinfo {author}
  {\bibfnamefont{F.}~\bibnamefont{Quevedo}},\ }%
  \bibfield{journal}{%
  \Doi{10.1007/JHEP10(2011)119}{\bibinfo {journal} {JHEP}}\ }%
  \textbf{\bibinfo {volume} {10}},\ \bibinfo {pages} {119} (\bibinfo {year}
  {2011}),\ \Eprint{http://arxiv.org/abs/1105.2107}{arXiv:1105.2107 [hep-th]}%
  \bibAnnoteFile{NoStop}{Cicoli:2011yy}%
%%CITATION = ARXIV:1105.2107;%%
\bibitem{Damour:1994zq}%
  \BibitemOpen
  \bibfield{author}{%
  \bibinfo {author} {\bibfnamefont{T.}~\bibnamefont{Damour}}\ and\ \bibinfo
  {author} {\bibfnamefont{A.~M.}\ \bibnamefont{Polyakov}},\ }%
  \bibfield{journal}{%
  \Doi{10.1016/0550-3213(94)90143-0}{\bibinfo {journal} {Nucl. Phys.}}\ }%
  \textbf{\bibinfo {volume} {B423}},\ \bibinfo {pages} {532} (\bibinfo {year}
  {1994}),\ \Eprint{http://arxiv.org/abs/hep-th/9401069}{arXiv:hep-th/9401069
  [hep-th]}%
  \bibAnnoteFile{NoStop}{Damour:1994zq}%
%%CITATION = HEP-TH/9401069;%%
\bibitem{Taylor:1988nw}%
  \BibitemOpen
  \bibfield{author}{%
  \bibinfo {author} {\bibfnamefont{T.~R.}\ \bibnamefont{Taylor}}\ and\ \bibinfo
  {author} {\bibfnamefont{G.}~\bibnamefont{Veneziano}},\ }%
  \bibfield{journal}{%
  \Doi{10.1016/0370-2693(88)91290-7}{\bibinfo {journal} {Phys. Lett.}}\ }%
  \textbf{\bibinfo {volume} {B213}},\ \bibinfo {pages} {450} (\bibinfo {year}
  {1988})%
  \bibAnnoteFile{NoStop}{Taylor:1988nw}%
%%CITATION = PHLTA,B213,450;%%
\bibitem{Piazza:2010ye}%
  \BibitemOpen
  \bibfield{author}{%
  \bibinfo {author} {\bibfnamefont{F.}~\bibnamefont{Piazza}}\ and\ \bibinfo
  {author} {\bibfnamefont{M.}~\bibnamefont{Pospelov}},\ }%
  \bibfield{journal}{%
  \Doi{10.1103/PhysRevD.82.043533}{\bibinfo {journal} {Phys. Rev.}}\ }%
  \textbf{\bibinfo {volume} {D82}},\ \bibinfo {pages} {043533} (\bibinfo {year}
  {2010}),\ \Eprint{http://arxiv.org/abs/1003.2313}{arXiv:1003.2313 [hep-ph]}%
  \bibAnnoteFile{NoStop}{Piazza:2010ye}%
%%CITATION = ARXIV:1003.2313;%%
\bibitem{Arvanitaki:2015iga}%
  \BibitemOpen
  \bibfield{author}{%
  \bibinfo {author} {\bibfnamefont{A.}~\bibnamefont{Arvanitaki}}, \bibinfo
  {author} {\bibfnamefont{S.}~\bibnamefont{Dimopoulos}},\ and\ \bibinfo
  {author} {\bibfnamefont{K.}~\bibnamefont{Van~Tilburg}},\ }%
  \bibfield{journal}{%
  \Doi{10.1103/PhysRevLett.116.031102}{\bibinfo {journal} {Phys. Rev. Lett.}}\
  }%
  \textbf{\bibinfo {volume} {116}},\ \bibinfo {pages} {031102} (\bibinfo {year}
  {2016}),\ \Eprint{http://arxiv.org/abs/1508.01798}{arXiv:1508.01798
  [hep-ph]}%
  \bibAnnoteFile{NoStop}{Arvanitaki:2015iga}%
%%CITATION = ARXIV:1508.01798;%%
\bibitem{Arvanitaki:2014faa}%
  \BibitemOpen
  \bibfield{author}{%
  \bibinfo {author} {\bibfnamefont{A.}~\bibnamefont{Arvanitaki}}, \bibinfo
  {author} {\bibfnamefont{J.}~\bibnamefont{Huang}},\ and\ \bibinfo {author}
  {\bibfnamefont{K.}~\bibnamefont{Van~Tilburg}},\ }%
  \bibfield{journal}{%
  \Doi{10.1103/PhysRevD.91.015015}{\bibinfo {journal} {Phys. Rev.}}\ }%
  \textbf{\bibinfo {volume} {D91}},\ \bibinfo {pages} {015015} (\bibinfo {year}
  {2015}),\ \Eprint{http://arxiv.org/abs/1405.2925}{arXiv:1405.2925 [hep-ph]}%
  \bibAnnoteFile{NoStop}{Arvanitaki:2014faa}%
%%CITATION = ARXIV:1405.2925;%%
\bibitem{Hees:2016gop}%
  \BibitemOpen
  \bibfield{author}{%
  \bibinfo {author} {\bibfnamefont{A.}~\bibnamefont{Hees}}, \bibinfo {author}
  {\bibfnamefont{J.}~\bibnamefont{GŽna}}, \bibinfo {author}
  {\bibfnamefont{M.}~\bibnamefont{Abgrall}}, \bibinfo {author}
  {\bibfnamefont{S.}~\bibnamefont{Bize}},\ and\ \bibinfo {author}
  {\bibfnamefont{P.}~\bibnamefont{Wolf}}}%
   (\bibinfo {year} {2016}),\
  \Eprint{http://arxiv.org/abs/1604.08514}{arXiv:1604.08514 [gr-qc]}%
  \bibAnnoteFile{NoStop}{Hees:2016gop}%
%%CITATION = ARXIV:1604.08514;%%
\bibitem{Graham:2015ifn}%
  \BibitemOpen
  \bibfield{author}{%
  \bibinfo {author} {\bibfnamefont{P.~W.}\ \bibnamefont{Graham}}, \bibinfo
  {author} {\bibfnamefont{D.~E.}\ \bibnamefont{Kaplan}}, \bibinfo {author}
  {\bibfnamefont{J.}~\bibnamefont{Mardon}}, \bibinfo {author}
  {\bibfnamefont{S.}~\bibnamefont{Rajendran}},\ and\ \bibinfo {author}
  {\bibfnamefont{W.~A.}\ \bibnamefont{Terrano}},\ }%
  \bibfield{journal}{%
  \Doi{10.1103/PhysRevD.93.075029}{\bibinfo {journal} {Phys. Rev.}}\ }%
  \textbf{\bibinfo {volume} {D93}},\ \bibinfo {pages} {075029} (\bibinfo {year}
  {2016}),\ \Eprint{http://arxiv.org/abs/1512.06165}{arXiv:1512.06165
  [hep-ph]}%
  \bibAnnoteFile{NoStop}{Graham:2015ifn}%
%%CITATION = ARXIV:1512.06165;%%
\bibitem{Stadnik:2015kia}%
  \BibitemOpen
  \bibfield{author}{%
  \bibinfo {author} {\bibfnamefont{Y.~V.}\ \bibnamefont{Stadnik}}\ and\
  \bibinfo {author} {\bibfnamefont{V.~V.}\ \bibnamefont{Flambaum}},\ }%
  \bibfield{journal}{%
  \Doi{10.1103/PhysRevLett.115.201301}{\bibinfo {journal} {Phys. Rev. Lett.}}\
  }%
  \textbf{\bibinfo {volume} {115}},\ \bibinfo {pages} {201301} (\bibinfo {year}
  {2015}),\ \Eprint{http://arxiv.org/abs/1503.08540}{arXiv:1503.08540
  [astro-ph.CO]}%
  \bibAnnoteFile{NoStop}{Stadnik:2015kia}%
%%CITATION = ARXIV:1503.08540;%%
\bibitem{Stadnik:2014tta}%
  \BibitemOpen
  \bibfield{author}{%
  \bibinfo {author} {\bibfnamefont{Y.~V.}\ \bibnamefont{Stadnik}}\ and\
  \bibinfo {author} {\bibfnamefont{V.~V.}\ \bibnamefont{Flambaum}},\ }%
  \bibfield{journal}{%
  \Doi{10.1103/PhysRevLett.114.161301}{\bibinfo {journal} {Phys. Rev. Lett.}}\
  }%
  \textbf{\bibinfo {volume} {114}},\ \bibinfo {pages} {161301} (\bibinfo {year}
  {2015}),\ \Eprint{http://arxiv.org/abs/1412.7801}{arXiv:1412.7801 [hep-ph]}%
  \bibAnnoteFile{NoStop}{Stadnik:2014tta}%
%%CITATION = ARXIV:1412.7801;%%
\bibitem{Arvanitaki:2016fyj}%
  \BibitemOpen
  \bibfield{author}{%
  \bibinfo {author} {\bibfnamefont{A.}~\bibnamefont{Arvanitaki}}, \bibinfo
  {author} {\bibfnamefont{P.~W.}\ \bibnamefont{Graham}}, \bibinfo {author}
  {\bibfnamefont{J.~M.}\ \bibnamefont{Hogan}}, \bibinfo {author}
  {\bibfnamefont{S.}~\bibnamefont{Rajendran}},\ and\ \bibinfo {author}
  {\bibfnamefont{K.}~\bibnamefont{Van~Tilburg}}}%
   (\bibinfo {year} {2016}),\
  \Eprint{http://arxiv.org/abs/1606.04541}{arXiv:1606.04541 [hep-ph]}%
  \bibAnnoteFile{NoStop}{Arvanitaki:2016fyj}%
%%CITATION = ARXIV:1606.04541;%%
\bibitem{Halprin:1997na}%
  \BibitemOpen
  \bibfield{author}{%
  \bibinfo {author} {\bibfnamefont{A.}~\bibnamefont{Halprin}}\ and\ \bibinfo
  {author} {\bibfnamefont{C.~N.}\ \bibnamefont{Leung}},\ }%
  \bibfield{journal}{%
  \Doi{10.1016/S0370-2693(97)01254-9}{\bibinfo {journal} {Phys. Lett.}}\ }%
  \textbf{\bibinfo {volume} {B416}},\ \bibinfo {pages} {361} (\bibinfo {year}
  {1998}),\ \Eprint{http://arxiv.org/abs/hep-ph/9707407}{arXiv:hep-ph/9707407
  [hep-ph]}%
  \bibAnnoteFile{NoStop}{Halprin:1997na}%
%%CITATION = HEP-PH/9707407;%%
\bibitem{Horvat:1998pp}%
  \BibitemOpen
  \bibfield{author}{%
  \bibinfo {author} {\bibfnamefont{R.}~\bibnamefont{Horvat}},\ }%
  \bibfield{journal}{%
  \Doi{10.1103/PhysRevD.58.125020}{\bibinfo {journal} {Phys. Rev.}}\ }%
  \textbf{\bibinfo {volume} {D58}},\ \bibinfo {pages} {125020} (\bibinfo {year}
  {1998}),\ \Eprint{http://arxiv.org/abs/hep-ph/9802377}{arXiv:hep-ph/9802377
  [hep-ph]}%
  \bibAnnoteFile{NoStop}{Horvat:1998pp}%
%%CITATION = HEP-PH/9802377;%%
\bibitem{Barger:2005mn}%
  \BibitemOpen
  \bibfield{author}{%
  \bibinfo {author} {\bibfnamefont{V.}~\bibnamefont{Barger}}, \bibinfo {author}
  {\bibfnamefont{P.}~\bibnamefont{Huber}},\ and\ \bibinfo {author}
  {\bibfnamefont{D.}~\bibnamefont{Marfatia}},\ }%
  \bibfield{journal}{%
  \Doi{10.1103/PhysRevLett.95.211802}{\bibinfo {journal} {Phys. Rev. Lett.}}\
  }%
  \textbf{\bibinfo {volume} {95}},\ \bibinfo {pages} {211802} (\bibinfo {year}
  {2005}),\ \Eprint{http://arxiv.org/abs/hep-ph/0502196}{arXiv:hep-ph/0502196
  [hep-ph]}%
  \bibAnnoteFile{NoStop}{Barger:2005mn}%
%%CITATION = HEP-PH/0502196;%%
\bibitem{Zurek:2004vd}%
  \BibitemOpen
  \bibfield{author}{%
  \bibinfo {author} {\bibfnamefont{K.~M.}\ \bibnamefont{Zurek}},\ }%
  \bibfield{journal}{%
  \Doi{10.1088/1126-6708/2004/10/058}{\bibinfo {journal} {JHEP}}\ }%
  \textbf{\bibinfo {volume} {10}},\ \bibinfo {pages} {058} (\bibinfo {year}
  {2004}),\ \Eprint{http://arxiv.org/abs/hep-ph/0405141}{arXiv:hep-ph/0405141
  [hep-ph]}%
  \bibAnnoteFile{NoStop}{Zurek:2004vd}%
%%CITATION = HEP-PH/0405141;%%
\bibitem{Kaplan:2004dq}%
  \BibitemOpen
  \bibfield{author}{%
  \bibinfo {author} {\bibfnamefont{D.~B.}\ \bibnamefont{Kaplan}}, \bibinfo
  {author} {\bibfnamefont{A.~E.}\ \bibnamefont{Nelson}},\ and\ \bibinfo
  {author} {\bibfnamefont{N.}~\bibnamefont{Weiner}},\ }%
  \bibfield{journal}{%
  \Doi{10.1103/PhysRevLett.93.091801}{\bibinfo {journal} {Phys. Rev. Lett.}}\
  }%
  \textbf{\bibinfo {volume} {93}},\ \bibinfo {pages} {091801} (\bibinfo {year}
  {2004}),\ \Eprint{http://arxiv.org/abs/hep-ph/0401099}{arXiv:hep-ph/0401099
  [hep-ph]}%
  \bibAnnoteFile{NoStop}{Kaplan:2004dq}%
%%CITATION = HEP-PH/0401099;%%
\bibitem{Barger:2005mh}%
  \BibitemOpen
  \bibfield{author}{%
  \bibinfo {author} {\bibfnamefont{V.}~\bibnamefont{Barger}}, \bibinfo {author}
  {\bibfnamefont{D.}~\bibnamefont{Marfatia}},\ and\ \bibinfo {author}
  {\bibfnamefont{K.}~\bibnamefont{Whisnant}},\ }%
  \bibfield{journal}{%
  \Doi{10.1103/PhysRevD.73.013005}{\bibinfo {journal} {Phys. Rev.}}\ }%
  \textbf{\bibinfo {volume} {D73}},\ \bibinfo {pages} {013005} (\bibinfo {year}
  {2006}),\ \Eprint{http://arxiv.org/abs/hep-ph/0509163}{arXiv:hep-ph/0509163
  [hep-ph]}%
  \bibAnnoteFile{NoStop}{Barger:2005mh}%
%%CITATION = HEP-PH/0509163;%%
\bibitem{Agashe:2014kda}%
  \BibitemOpen
  \bibfield{author}{%
  \bibinfo {author} {\bibfnamefont{K.~A.}\ \bibnamefont{Olive}} \emph{et~al.}
  (\bibinfo {collaboration} {Particle Data Group}),\ }%
  \bibfield{journal}{%
  \Doi{10.1088/1674-1137/38/9/090001}{\bibinfo {journal} {Chin. Phys.}}\ }%
  \textbf{\bibinfo {volume} {C38}},\ \bibinfo {pages} {090001} (\bibinfo {year}
  {2014})%
  \bibAnnoteFile{NoStop}{Agashe:2014kda}%
%%CITATION = CHPHD,C38,090001;%%
\bibitem{Willenbrock:2004hu}%
  \BibitemOpen
  \bibfield{author}{%
  \bibinfo {author} {\bibfnamefont{S.}~\bibnamefont{Willenbrock}},\ }%
  in\ \emph{\bibinfo {booktitle} {{Physics in D >= 4. Proceedings, Theoretical
  Advanced Study Institute in elementary particle physics, TASI 2004, Boulder,
  USA, June 6-July 2, 2004}}}\ (\bibinfo {year} {2004})\ pp.\ \bibinfo {pages}
  {3--38},\ \Eprint{http://arxiv.org/abs/hep-ph/0410370}{arXiv:hep-ph/0410370
  [hep-ph]}%
  \bibAnnoteFile{NoStop}{Willenbrock:2004hu}%
%%CITATION = HEP-PH/0410370;%%
\bibitem{Schechter:1980gr}%
  \BibitemOpen
  \bibfield{author}{%
  \bibinfo {author} {\bibfnamefont{J.}~\bibnamefont{Schechter}}\ and\ \bibinfo
  {author} {\bibfnamefont{J.~W.~F.}\ \bibnamefont{Valle}},\ }%
  \bibfield{journal}{%
  \Doi{10.1103/PhysRevD.22.2227}{\bibinfo {journal} {Phys. Rev.}}\ }%
  \textbf{\bibinfo {volume} {D22}},\ \bibinfo {pages} {2227} (\bibinfo {year}
  {1980})%
  \bibAnnoteFile{NoStop}{Schechter:1980gr}%
%%CITATION = PHRVA,D22,2227;%%
\bibitem{GellMann:1980vs}%
  \BibitemOpen
  \bibfield{author}{%
  \bibinfo {author} {\bibfnamefont{M.}~\bibnamefont{Gell-Mann}}, \bibinfo
  {author} {\bibfnamefont{P.}~\bibnamefont{Ramond}},\ and\ \bibinfo {author}
  {\bibfnamefont{R.}~\bibnamefont{Slansky}},\ }%
  \bibfield{booktitle}{%
  \emph{\bibinfo {booktitle} {{Supergravity Workshop Stony Brook, New York,
  September 27-28, 1979}}},\ }%
  \bibfield{journal}{%
  \bibinfo {journal} {Conf. Proc.}\ }%
  \textbf{\bibinfo {volume} {C790927}},\ \bibinfo {pages} {315} (\bibinfo
  {year} {1979}),\ \Eprint{http://arxiv.org/abs/1306.4669}{arXiv:1306.4669
  [hep-th]}%
  \bibAnnoteFile{NoStop}{GellMann:1980vs}%
%%CITATION = ARXIV:1306.4669;%%
\bibitem{Mohapatra:1979ia}%
  \BibitemOpen
  \bibfield{author}{%
  \bibinfo {author} {\bibfnamefont{R.~N.}\ \bibnamefont{Mohapatra}}\ and\
  \bibinfo {author} {\bibfnamefont{G.}~\bibnamefont{Senjanovic}},\ }%
  \bibfield{journal}{%
  \Doi{10.1103/PhysRevLett.44.912}{\bibinfo {journal} {Phys. Rev. Lett.}}\ }%
  \textbf{\bibinfo {volume} {44}},\ \bibinfo {pages} {912} (\bibinfo {year}
  {1980})%
  \bibAnnoteFile{NoStop}{Mohapatra:1979ia}%
%%CITATION = PRLTA,44,912;%%
\bibitem{Yanagida:1979as}%
  \BibitemOpen
  \bibfield{author}{%
  \bibinfo {author} {\bibfnamefont{T.}~\bibnamefont{Yanagida}},\ }%
  \bibfield{booktitle}{%
  \emph{\bibinfo {booktitle} {{Proceedings: Workshop on the Unified Theories
  and the Baryon Number in the Universe, Tsukuba, Japan, 13-14 Feb 1979}}},\ }%
  \bibfield{journal}{%
  \bibinfo {journal} {Conf. Proc.}\ }%
  \textbf{\bibinfo {volume} {C7902131}},\ \bibinfo {pages} {95} (\bibinfo
  {year} {1979})%
  \bibAnnoteFile{NoStop}{Yanagida:1979as}%
%%CITATION = CONFP,C7902131,95;%%
\bibitem{Minkowski:1977sc}%
  \BibitemOpen
  \bibfield{author}{%
  \bibinfo {author} {\bibfnamefont{P.}~\bibnamefont{Minkowski}},\ }%
  \bibfield{journal}{%
  \Doi{10.1016/0370-2693(77)90435-X}{\bibinfo {journal} {Phys. Lett.}}\ }%
  \textbf{\bibinfo {volume} {B67}},\ \bibinfo {pages} {421} (\bibinfo {year}
  {1977})%
  \bibAnnoteFile{NoStop}{Minkowski:1977sc}%
%%CITATION = PHLTA,B67,421;%%
\bibitem{Fukuda:2001nj}%
  \BibitemOpen
  \bibfield{author}{%
  \bibinfo {author} {\bibfnamefont{S.}~\bibnamefont{Fukuda}} \emph{et~al.}
  (\bibinfo {collaboration} {Super-Kamiokande}),\ }%
  \bibfield{journal}{%
  \Doi{10.1103/PhysRevLett.86.5651}{\bibinfo {journal} {Phys. Rev. Lett.}}\ }%
  \textbf{\bibinfo {volume} {86}},\ \bibinfo {pages} {5651} (\bibinfo {year}
  {2001}),\ \Eprint{http://arxiv.org/abs/hep-ex/0103032}{arXiv:hep-ex/0103032
  [hep-ex]}%
  \bibAnnoteFile{NoStop}{Fukuda:2001nj}%
%%CITATION = HEP-EX/0103032;%%
\bibitem{deGouvea:1999wg}%
  \BibitemOpen
  \bibfield{author}{%
  \bibinfo {author} {\bibfnamefont{A.}~\bibnamefont{de~Gouvea}}, \bibinfo
  {author} {\bibfnamefont{A.}~\bibnamefont{Friedland}},\ and\ \bibinfo {author}
  {\bibfnamefont{H.}~\bibnamefont{Murayama}},\ }%
  \bibfield{journal}{%
  \Doi{10.1103/PhysRevD.60.093011}{\bibinfo {journal} {Phys. Rev.}}\ }%
  \textbf{\bibinfo {volume} {D60}},\ \bibinfo {pages} {093011} (\bibinfo {year}
  {1999}),\ \Eprint{http://arxiv.org/abs/hep-ph/9904399}{arXiv:hep-ph/9904399
  [hep-ph]}%
  \bibAnnoteFile{NoStop}{deGouvea:1999wg}%
%%CITATION = HEP-PH/9904399;%%
\bibitem{Gonzalez-Garcia:2014bfa}%
  \BibitemOpen
  \bibfield{author}{%
  \bibinfo {author} {\bibfnamefont{M.~C.}\ \bibnamefont{Gonzalez-Garcia}},
  \bibinfo {author} {\bibfnamefont{M.}~\bibnamefont{Maltoni}},\ and\ \bibinfo
  {author} {\bibfnamefont{T.}~\bibnamefont{Schwetz}},\ }%
  \bibfield{journal}{%
  \Doi{10.1007/JHEP11(2014)052}{\bibinfo {journal} {JHEP}}\ }%
  \textbf{\bibinfo {volume} {11}},\ \bibinfo {pages} {052} (\bibinfo {year}
  {2014}),\ \Eprint{http://arxiv.org/abs/1409.5439}{arXiv:1409.5439 [hep-ph]}%
  \bibAnnoteFile{NoStop}{Gonzalez-Garcia:2014bfa}%
%%CITATION = ARXIV:1409.5439;%%
\bibitem{Bahcall:1990dr}%
  \BibitemOpen
  \bibfield{author}{%
  \bibinfo {author} {\bibfnamefont{J.~N.}\ \bibnamefont{Bahcall}}\ and\
  \bibinfo {author} {\bibfnamefont{W.~H.}\ \bibnamefont{Press}},\ }%
  \bibfield{journal}{%
  \Doi{10.1086/169856}{\bibinfo {journal} {Astrophys. J.}}\ }%
  \textbf{\bibinfo {volume} {370}},\ \bibinfo {pages} {730} (\bibinfo {year}
  {1991})%
  \bibAnnoteFile{NoStop}{Bahcall:1990dr}%
%%CITATION = ASJOA,370,730;%%
\bibitem{Sturrock:1997gp}%
  \BibitemOpen
  \bibfield{author}{%
  \bibinfo {author} {\bibfnamefont{P.~A.}\ \bibnamefont{Sturrock}}, \bibinfo
  {author} {\bibfnamefont{G.}~\bibnamefont{Walther}},\ and\ \bibinfo {author}
  {\bibfnamefont{M.~S.}\ \bibnamefont{Wheatland}},\ }%
  \bibfield{journal}{%
  \Doi{10.1086/304955}{\bibinfo {journal} {Astrophys. J.}}\ }%
  \textbf{\bibinfo {volume} {491}},\ \bibinfo {pages} {409} (\bibinfo {year}
  {1997})%
  \bibAnnoteFile{NoStop}{Sturrock:1997gp}%
%%CITATION = ASJOA,491,409;%%
\bibitem{Caldwell:2003dw}%
  \BibitemOpen
  \bibfield{author}{%
  \bibinfo {author} {\bibfnamefont{D.~O.}\ \bibnamefont{Caldwell}}\ and\
  \bibinfo {author} {\bibfnamefont{P.~A.}\ \bibnamefont{Sturrock}},\ }%
  \bibfield{journal}{%
  \Doi{10.1016/j.astropartphys.2005.05.001}{\bibinfo {journal} {Astropart.
  Phys.}}\ }%
  \textbf{\bibinfo {volume} {23}},\ \bibinfo {pages} {543} (\bibinfo {year}
  {2005}),\ \Eprint{http://arxiv.org/abs/hep-ph/0309191}{arXiv:hep-ph/0309191
  [hep-ph]}%
  \bibAnnoteFile{NoStop}{Caldwell:2003dw}%
%%CITATION = HEP-PH/0309191;%%
\bibitem{Milsztajn:2003af}%
  \BibitemOpen
  \bibfield{author}{%
  \bibinfo {author} {\bibfnamefont{A.}~\bibnamefont{Milsztajn}}}%
   (\bibinfo {year} {2003}),\
  \Eprint{http://arxiv.org/abs/hep-ph/0301252}{arXiv:hep-ph/0301252 [hep-ph]}%
  \bibAnnoteFile{NoStop}{Milsztajn:2003af}%
%%CITATION = HEP-PH/0301252;%%
\bibitem{Ghosh:2006me}%
  \BibitemOpen
  \bibfield{author}{%
  \bibinfo {author} {\bibfnamefont{K.}~\bibnamefont{Ghosh}}\ and\ \bibinfo
  {author} {\bibfnamefont{P.}~\bibnamefont{Raychaudhuri}}}%
   (\bibinfo {year} {2006}),\
  \Eprint{http://arxiv.org/abs/hep-ph/0606317}{arXiv:hep-ph/0606317 [hep-ph]}%
  \bibAnnoteFile{NoStop}{Ghosh:2006me}%
%%CITATION = HEP-PH/0606317;%%
\bibitem{Ghosh:2006xk}%
  \BibitemOpen
  \bibfield{author}{%
  \bibinfo {author} {\bibfnamefont{K.}~\bibnamefont{Ghosh}}\ and\ \bibinfo
  {author} {\bibfnamefont{P.}~\bibnamefont{Raychaudhuri}},\ }%
  in\ \emph{\bibinfo {booktitle} {{29th International Cosmic Ray Conference
  (ICRC 2005) Pune, India, August 3-11, 2005}}}\ (\bibinfo {year} {2006})\
  \Eprint{http://arxiv.org/abs/astro-ph/0606083}{arXiv:astro-ph/0606083
  [astro-ph]}%
  \bibAnnoteFile{NoStop}{Ghosh:2006xk}%
%%CITATION = ASTRO-PH/0606083;%%
\bibitem{Sturrock:2003ep}%
  \BibitemOpen
  \bibfield{author}{%
  \bibinfo {author} {\bibfnamefont{P.~A.}\ \bibnamefont{Sturrock}},\ }%
  \bibfield{journal}{%
  \Doi{10.1086/377079}{\bibinfo {journal} {Astrophys. J.}}\ }%
  \textbf{\bibinfo {volume} {594}},\ \bibinfo {pages} {1102} (\bibinfo {year}
  {2003}),\ \Eprint{http://arxiv.org/abs/hep-ph/0304073}{arXiv:hep-ph/0304073
  [hep-ph]}%
  \bibAnnoteFile{NoStop}{Sturrock:2003ep}%
%%CITATION = HEP-PH/0304073;%%
\bibitem{Sturrock:2000jk}%
  \BibitemOpen
  \bibfield{author}{%
  \bibinfo {author} {\bibfnamefont{P.~A.}\ \bibnamefont{Sturrock}}\ and\
  \bibinfo {author} {\bibfnamefont{J.~D.}\ \bibnamefont{Scargle}},\ }%
  \bibfield{journal}{%
  \Doi{10.1086/319482}{\bibinfo {journal} {Astrophys. J.}}\ }%
  \textbf{\bibinfo {volume} {550}},\ \bibinfo {pages} {L101} (\bibinfo {year}
  {2001}),\
  \Eprint{http://arxiv.org/abs/astro-ph/0011228}{arXiv:astro-ph/0011228
  [astro-ph]}%
  \bibAnnoteFile{NoStop}{Sturrock:2000jk}%
%%CITATION = ASTRO-PH/0011228;%%
\bibitem{Sturrock:2003kv}%
  \BibitemOpen
  \bibfield{author}{%
  \bibinfo {author} {\bibfnamefont{P.~A.}\ \bibnamefont{Sturrock}},\ }%
  \bibfield{journal}{%
  \Doi{10.1086/382141}{\bibinfo {journal} {Astrophys. J.}}\ }%
  \textbf{\bibinfo {volume} {605}},\ \bibinfo {pages} {568} (\bibinfo {year}
  {2004}),\ \Eprint{http://arxiv.org/abs/hep-ph/0309239}{arXiv:hep-ph/0309239
  [hep-ph]}%
  \bibAnnoteFile{NoStop}{Sturrock:2003kv}%
%%CITATION = HEP-PH/0309239;%%
\bibitem{Sturrock:2005wf}%
  \BibitemOpen
  \bibfield{author}{%
  \bibinfo {author} {\bibfnamefont{P.~A.}\ \bibnamefont{Sturrock}}, \bibinfo
  {author} {\bibfnamefont{D.~O.}\ \bibnamefont{Caldwell}}, \bibinfo {author}
  {\bibfnamefont{J.~D.}\ \bibnamefont{Scargle}},\ and\ \bibinfo {author}
  {\bibfnamefont{M.~S.}\ \bibnamefont{Wheatland}},\ }%
  \bibfield{journal}{%
  \Doi{10.1103/PhysRevD.72.113004}{\bibinfo {journal} {Phys. Rev.}}\ }%
  \textbf{\bibinfo {volume} {D72}},\ \bibinfo {pages} {113004} (\bibinfo {year}
  {2005}),\ \Eprint{http://arxiv.org/abs/hep-ph/0501205}{arXiv:hep-ph/0501205
  [hep-ph]}%
  \bibAnnoteFile{NoStop}{Sturrock:2005wf}%
%%CITATION = HEP-PH/0501205;%%
\bibitem{Sturrock:2004hx}%
  \BibitemOpen
  \bibfield{author}{%
  \bibinfo {author} {\bibfnamefont{P.~A.}\ \bibnamefont{Sturrock}}, \bibinfo
  {author} {\bibfnamefont{D.~O.}\ \bibnamefont{Caldwell}}, \bibinfo {author}
  {\bibfnamefont{J.~D.}\ \bibnamefont{Scargle}}, \bibinfo {author}
  {\bibfnamefont{G.}~\bibnamefont{Walther}},\ and\ \bibinfo {author}
  {\bibfnamefont{M.~S.}\ \bibnamefont{Wheatland}}}%
   (\bibinfo {year} {2004}),\
  \Eprint{http://arxiv.org/abs/hep-ph/0403246}{arXiv:hep-ph/0403246 [hep-ph]}%
  \bibAnnoteFile{NoStop}{Sturrock:2004hx}%
%%CITATION = HEP-PH/0403246;%%
\bibitem{Sturrock:2004jv}%
  \BibitemOpen
  \bibfield{author}{%
  \bibinfo {author} {\bibfnamefont{P.~A.}\ \bibnamefont{Sturrock}}\ and\
  \bibinfo {author} {\bibfnamefont{D.~O.}\ \bibnamefont{Caldwell}},\ }%
  \bibfield{journal}{%
  \Doi{10.1016/j.astropartphys.2006.06.001}{\bibinfo {journal} {Astropart.
  Phys.}}\ }%
  \textbf{\bibinfo {volume} {26}},\ \bibinfo {pages} {174} (\bibinfo {year}
  {2006}),\ \Eprint{http://arxiv.org/abs/hep-ph/0409064}{arXiv:hep-ph/0409064
  [hep-ph]}%
  \bibAnnoteFile{NoStop}{Sturrock:2004jv}%
%%CITATION = HEP-PH/0409064;%%
\bibitem{Sturrock:2006qz}%
  \BibitemOpen
  \bibfield{author}{%
  \bibinfo {author} {\bibfnamefont{P.~A.}\ \bibnamefont{Sturrock}},\ }%
  \bibfield{journal}{%
  \Doi{10.1007/s11207-006-0143-0}{\bibinfo {journal} {Solar Phys.}}\ }%
  \textbf{\bibinfo {volume} {237}},\ \bibinfo {pages} {1} (\bibinfo {year}
  {2006}),\ \Eprint{http://arxiv.org/abs/hep-ph/0601251}{arXiv:hep-ph/0601251
  [hep-ph]}%
  \bibAnnoteFile{NoStop}{Sturrock:2006qz}%
%%CITATION = HEP-PH/0601251;%%
\bibitem{Sturrock:2009cm}%
  \BibitemOpen
  \bibfield{author}{%
  \bibinfo {author} {\bibfnamefont{P.~A.}\ \bibnamefont{Sturrock}},\ }%
  \bibfield{journal}{%
  \Doi{10.1007/s11207-009-9462-2}{\bibinfo {journal} {Solar Phys.}}\ }%
  \textbf{\bibinfo {volume} {260}},\ \bibinfo {pages} {245} (\bibinfo {year}
  {2009}),\ \Eprint{http://arxiv.org/abs/0904.4236}{arXiv:0904.4236
  [astro-ph.HE]}%
  \bibAnnoteFile{NoStop}{Sturrock:2009cm}%
%%CITATION = ARXIV:0904.4236;%%
\bibitem{Sturrock:2012re}%
  \BibitemOpen
  \bibfield{author}{%
  \bibinfo {author} {\bibfnamefont{P.~A.}\ \bibnamefont{Sturrock}}, \bibinfo
  {author} {\bibfnamefont{L.}~\bibnamefont{Bertello}}, \bibinfo {author}
  {\bibfnamefont{E.}~\bibnamefont{Fischbach}}, \bibinfo {author}
  {\bibfnamefont{D.}~\bibnamefont{Javorsek}, \bibfnamefont{II}}, \bibinfo
  {author} {\bibfnamefont{J.~H.}\ \bibnamefont{Jenkins}}, \bibinfo {author}
  {\bibfnamefont{A.}~\bibnamefont{Kosovichev}},\ and\ \bibinfo {author}
  {\bibfnamefont{A.~G.}\ \bibnamefont{Parkhomov}},\ }%
  \bibfield{journal}{%
  \Doi{10.1016/j.astropartphys.2012.11.011}{\bibinfo {journal} {Astropart.
  Phys.}}\ }%
  \textbf{\bibinfo {volume} {42}},\ \bibinfo {pages} {62} (\bibinfo {year}
  {2013}),\ \Eprint{http://arxiv.org/abs/1211.6352}{arXiv:1211.6352
  [astro-ph.SR]}%
  \bibAnnoteFile{NoStop}{Sturrock:2012re}%
%%CITATION = ARXIV:1211.6352;%%
\bibitem{Sturrock:2015ivo}%
  \BibitemOpen
  \bibfield{author}{%
  \bibinfo {author} {\bibfnamefont{P.~A.}\ \bibnamefont{Sturrock}}\ and\
  \bibinfo {author} {\bibfnamefont{E.}~\bibnamefont{Fischbach}}}%
   (\bibinfo {year} {2015}),\
  \Eprint{http://arxiv.org/abs/1511.08770}{arXiv:1511.08770 [hep-ph]}%
  \bibAnnoteFile{NoStop}{Sturrock:2015ivo}%
%%CITATION = ARXIV:1511.08770;%%
\bibitem{Desai:2016bmz}%
  \BibitemOpen
  \bibfield{author}{%
  \bibinfo {author} {\bibfnamefont{S.}~\bibnamefont{Desai}}\ and\ \bibinfo
  {author} {\bibfnamefont{D.~W.}\ \bibnamefont{Liu}},\ }%
  \bibfield{journal}{%
  \Doi{10.1016/j.astropartphys.2016.06.004}{\bibinfo {journal} {Astropart.
  Phys.}}\ }%
  \textbf{\bibinfo {volume} {82}},\ \bibinfo {pages} {86} (\bibinfo {year}
  {2016}),\ \Eprint{http://arxiv.org/abs/1604.06758}{arXiv:1604.06758
  [astro-ph.HE]}%
  \bibAnnoteFile{NoStop}{Desai:2016bmz}%
%%CITATION = ARXIV:1604.06758;%%
\bibitem{Sturrock:2008dg}%
  \BibitemOpen
  \bibfield{author}{%
  \bibinfo {author} {\bibfnamefont{P.~A.}\ \bibnamefont{Sturrock}},\ }%
  \bibfield{journal}{%
  \Doi{10.1007/s11207-008-9253-1}{\bibinfo {journal} {Solar Phys.}}\ }%
  \textbf{\bibinfo {volume} {252}},\ \bibinfo {pages} {221} (\bibinfo {year}
  {2008}),\ \Eprint{http://arxiv.org/abs/0802.3399}{arXiv:0802.3399 [hep-ph]}%
  \bibAnnoteFile{NoStop}{Sturrock:2008dg}%
%%CITATION = ARXIV:0802.3399;%%
\bibitem{Akhmedov:2002mf}%
  \BibitemOpen
  \bibfield{author}{%
  \bibinfo {author} {\bibfnamefont{E.~K.}\ \bibnamefont{Akhmedov}}\ and\
  \bibinfo {author} {\bibfnamefont{J.}~\bibnamefont{Pulido}},\ }%
  \bibfield{journal}{%
  \Doi{10.1016/S0370-2693(02)03182-9}{\bibinfo {journal} {Phys. Lett.}}\ }%
  \textbf{\bibinfo {volume} {B553}},\ \bibinfo {pages} {7} (\bibinfo {year}
  {2003}),\ \Eprint{http://arxiv.org/abs/hep-ph/0209192}{arXiv:hep-ph/0209192
  [hep-ph]}%
  \bibAnnoteFile{NoStop}{Akhmedov:2002mf}%
%%CITATION = HEP-PH/0209192;%%
\bibitem{Ranucci:2007fb}%
  \BibitemOpen
  \bibfield{author}{%
  \bibinfo {author} {\bibfnamefont{G.}~\bibnamefont{Ranucci}}\ and\ \bibinfo
  {author} {\bibfnamefont{S.}~\bibnamefont{Sello}},\ }%
  \bibfield{journal}{%
  \Doi{10.1103/PhysRevD.75.073011}{\bibinfo {journal} {Phys. Rev.}}\ }%
  \textbf{\bibinfo {volume} {D75}},\ \bibinfo {pages} {073011} (\bibinfo {year}
  {2007})%
  \bibAnnoteFile{NoStop}{Ranucci:2007fb}%
%%CITATION = PHRVA,D75,073011;%%
\bibitem{Cravens:2008aa}%
  \BibitemOpen
  \bibfield{author}{%
  \bibinfo {author} {\bibfnamefont{J.~P.}\ \bibnamefont{Cravens}} \emph{et~al.}
  (\bibinfo {collaboration} {Super-Kamiokande}),\ }%
  \bibfield{journal}{%
  \Doi{10.1103/PhysRevD.78.032002}{\bibinfo {journal} {Phys. Rev.}}\ }%
  \textbf{\bibinfo {volume} {D78}},\ \bibinfo {pages} {032002} (\bibinfo {year}
  {2008}),\ \Eprint{http://arxiv.org/abs/0803.4312}{arXiv:0803.4312 [hep-ex]}%
  \bibAnnoteFile{NoStop}{Cravens:2008aa}%
%%CITATION = ARXIV:0803.4312;%%
\bibitem{Yoo:2003rc}%
  \BibitemOpen
  \bibfield{author}{%
  \bibinfo {author} {\bibfnamefont{J.}~\bibnamefont{Yoo}} \emph{et~al.}
  (\bibinfo {collaboration} {Super-Kamiokande}),\ }%
  \bibfield{journal}{%
  \Doi{10.1103/PhysRevD.68.092002}{\bibinfo {journal} {Phys. Rev.}}\ }%
  \textbf{\bibinfo {volume} {D68}},\ \bibinfo {pages} {092002} (\bibinfo {year}
  {2003}),\ \Eprint{http://arxiv.org/abs/hep-ex/0307070}{arXiv:hep-ex/0307070
  [hep-ex]}%
  \bibAnnoteFile{NoStop}{Yoo:2003rc}%
%%CITATION = HEP-EX/0307070;%%
\bibitem{Aharmim:2005iu}%
  \BibitemOpen
  \bibfield{author}{%
  \bibinfo {author} {\bibfnamefont{B.}~\bibnamefont{Aharmim}} \emph{et~al.}
  (\bibinfo {collaboration} {SNO}),\ }%
  \bibfield{journal}{%
  \Doi{10.1103/PhysRevD.72.052010}{\bibinfo {journal} {Phys. Rev.}}\ }%
  \textbf{\bibinfo {volume} {D72}},\ \bibinfo {pages} {052010} (\bibinfo {year}
  {2005}),\ \Eprint{http://arxiv.org/abs/hep-ex/0507079}{arXiv:hep-ex/0507079
  [hep-ex]}%
  \bibAnnoteFile{NoStop}{Aharmim:2005iu}%
%%CITATION = HEP-EX/0507079;%%
\bibitem{Collaboration:2009qz}%
  \BibitemOpen
  \bibfield{author}{%
  \bibinfo {author} {\bibfnamefont{B.}~\bibnamefont{Aharmim}} \emph{et~al.}
  (\bibinfo {collaboration} {SNO}),\ }%
  \bibfield{journal}{%
  \Doi{10.1088/0004-637X/710/1/540}{\bibinfo {journal} {Astrophys. J.}}\ }%
  \textbf{\bibinfo {volume} {710}},\ \bibinfo {pages} {540} (\bibinfo {year}
  {2010}),\ \Eprint{http://arxiv.org/abs/0910.2433}{arXiv:0910.2433
  [astro-ph.SR]}%
  \bibAnnoteFile{NoStop}{Collaboration:2009qz}%
%%CITATION = ARXIV:0910.2433;%%
\bibitem{Wurm:2010mq}%
  \BibitemOpen
  \bibfield{author}{%
  \bibinfo {author} {\bibfnamefont{M.}~\bibnamefont{Wurm}} \emph{et~al.},\ }%
  \bibfield{journal}{%
  \Doi{10.1103/PhysRevD.83.032010}{\bibinfo {journal} {Phys. Rev.}}\ }%
  \textbf{\bibinfo {volume} {D83}},\ \bibinfo {pages} {032010} (\bibinfo {year}
  {2011}),\ \Eprint{http://arxiv.org/abs/1012.3021}{arXiv:1012.3021
  [astro-ph.IM]}%
  \bibAnnoteFile{NoStop}{Wurm:2010mq}%
%%CITATION = ARXIV:1012.3021;%%
\bibitem{Alenazi:2006wu}%
  \BibitemOpen
  \bibfield{author}{%
  \bibinfo {author} {\bibfnamefont{M.~S.}\ \bibnamefont{Alenazi}}\ and\
  \bibinfo {author} {\bibfnamefont{P.}~\bibnamefont{Gondolo}},\ }%
  \bibfield{journal}{%
  \Doi{10.1103/PhysRevD.74.083518}{\bibinfo {journal} {Phys. Rev.}}\ }%
  \textbf{\bibinfo {volume} {D74}},\ \bibinfo {pages} {083518} (\bibinfo {year}
  {2006}),\
  \Eprint{http://arxiv.org/abs/astro-ph/0608390}{arXiv:astro-ph/0608390
  [astro-ph]}%
  \bibAnnoteFile{NoStop}{Alenazi:2006wu}%
%%CITATION = ASTRO-PH/0608390;%%
\bibitem{Bromley:2011aa}%
  \BibitemOpen
  \bibfield{author}{%
  \bibinfo {author} {\bibfnamefont{B.~C.}\ \bibnamefont{Bromley}},\ }%
  \bibfield{journal}{%
  \Doi{10.1088/0067-0049/197/2/37}{\bibinfo {journal} {Astrophys. J. Suppl.}}\
  }%
  \textbf{\bibinfo {volume} {197}},\ \bibinfo {pages} {37} (\bibinfo {year}
  {2011}),\ \Eprint{http://arxiv.org/abs/1112.2355}{arXiv:1112.2355
  [astro-ph.HE]}%
  \bibAnnoteFile{NoStop}{Bromley:2011aa}%
%%CITATION = ARXIV:1112.2355;%%
\bibitem{Pasquini:2015fjv}%
  \BibitemOpen
  \bibfield{author}{%
  \bibinfo {author} {\bibfnamefont{P.~S.}\ \bibnamefont{Pasquini}}\ and\
  \bibinfo {author} {\bibfnamefont{O.~L.~G.}\ \bibnamefont{Peres}},\ }%
  \bibfield{journal}{%
  \Doi{10.1103/PhysRevD.93.053007, 10.1103/PhysRevD.93.079902}{\bibinfo
  {journal} {Phys. Rev.}}\ }%
  \textbf{\bibinfo {volume} {D93}},\ \bibinfo {pages} {053007} (\bibinfo {year}
  {2016}),\ \bibinfo {note} {[Erratum: Phys. Rev.D93,no.7,079902(2016)]},\
  \Eprint{http://arxiv.org/abs/1511.01811}{arXiv:1511.01811 [hep-ph]}%
  \bibAnnoteFile{NoStop}{Pasquini:2015fjv}%
%%CITATION = ARXIV:1511.01811;%%
\bibitem{Abazajian:2011dt}%
  \BibitemOpen
  \bibfield{author}{%
  \bibinfo {author} {\bibfnamefont{K.~N.}\ \bibnamefont{Abazajian}}
  \emph{et~al.},\ }%
  \bibfield{journal}{%
  \Doi{10.1016/j.astropartphys.2011.07.002}{\bibinfo {journal} {Astropart.
  Phys.}}\ }%
  \textbf{\bibinfo {volume} {35}},\ \bibinfo {pages} {177} (\bibinfo {year}
  {2011}),\ \Eprint{http://arxiv.org/abs/1103.5083}{arXiv:1103.5083
  [astro-ph.CO]}%
  \bibAnnoteFile{NoStop}{Abazajian:2011dt}%
%%CITATION = ARXIV:1103.5083;%%
\bibitem{Ade:2015xua}%
  \BibitemOpen
  \bibfield{author}{%
  \bibinfo {author} {\bibfnamefont{P.~A.~R.}\ \bibnamefont{Ade}} \emph{et~al.}
  (\bibinfo {collaboration} {Planck})}%
   (\bibinfo {year} {2015}),\
  \Eprint{http://arxiv.org/abs/1502.01589}{arXiv:1502.01589 [astro-ph.CO]}%
  \bibAnnoteFile{NoStop}{Ade:2015xua}%
%%CITATION = ARXIV:1502.01589;%%
\bibitem{Wu:2014hta}%
  \BibitemOpen
  \bibfield{author}{%
  \bibinfo {author} {\bibfnamefont{W.~L.~K.}\ \bibnamefont{Wu}}, \bibinfo
  {author} {\bibfnamefont{J.}~\bibnamefont{Errard}}, \bibinfo {author}
  {\bibfnamefont{C.}~\bibnamefont{Dvorkin}}, \bibinfo {author}
  {\bibfnamefont{C.~L.}\ \bibnamefont{Kuo}}, \bibinfo {author}
  {\bibfnamefont{A.~T.}\ \bibnamefont{Lee}}, \bibinfo {author}
  {\bibfnamefont{P.}~\bibnamefont{McDonald}}, \bibinfo {author}
  {\bibfnamefont{A.}~\bibnamefont{Slosar}},\ and\ \bibinfo {author}
  {\bibfnamefont{O.}~\bibnamefont{Zahn}},\ }%
  \bibfield{journal}{%
  \Doi{10.1088/0004-637X/788/2/138}{\bibinfo {journal} {Astrophys. J.}}\ }%
  \textbf{\bibinfo {volume} {788}},\ \bibinfo {pages} {138} (\bibinfo {year}
  {2014}),\ \Eprint{http://arxiv.org/abs/1402.4108}{arXiv:1402.4108
  [astro-ph.CO]}%
  \bibAnnoteFile{NoStop}{Wu:2014hta}%
%%CITATION = ARXIV:1402.4108;%%
\bibitem{Mangano:2011ar}%
  \BibitemOpen
  \bibfield{author}{%
  \bibinfo {author} {\bibfnamefont{G.}~\bibnamefont{Mangano}}\ and\ \bibinfo
  {author} {\bibfnamefont{P.~D.}\ \bibnamefont{Serpico}},\ }%
  \bibfield{journal}{%
  \Doi{10.1016/j.physletb.2011.05.075}{\bibinfo {journal} {Phys. Lett.}}\ }%
  \textbf{\bibinfo {volume} {B701}},\ \bibinfo {pages} {296} (\bibinfo {year}
  {2011}),\ \Eprint{http://arxiv.org/abs/1103.1261}{arXiv:1103.1261
  [astro-ph.CO]}%
  \bibAnnoteFile{NoStop}{Mangano:2011ar}%
%%CITATION = ARXIV:1103.1261;%%
\bibitem{Hui:2016ltb}%
  \BibitemOpen
  \bibfield{author}{%
  \bibinfo {author} {\bibfnamefont{L.}~\bibnamefont{Hui}}, \bibinfo {author}
  {\bibfnamefont{J.~P.}\ \bibnamefont{Ostriker}}, \bibinfo {author}
  {\bibfnamefont{S.}~\bibnamefont{Tremaine}},\ and\ \bibinfo {author}
  {\bibfnamefont{E.}~\bibnamefont{Witten}}}%
   (\bibinfo {year} {2016}),\
  \Eprint{http://arxiv.org/abs/1610.08297}{arXiv:1610.08297 [astro-ph.CO]}%
  \bibAnnoteFile{NoStop}{Hui:2016ltb}%
%%CITATION = ARXIV:1610.08297;%%
\bibitem{Fogli:2005qa}%
  \BibitemOpen
  \bibfield{author}{%
  \bibinfo {author} {\bibfnamefont{G.~L.}\ \bibnamefont{Fogli}}, \bibinfo
  {author} {\bibfnamefont{E.}~\bibnamefont{Lisi}}, \bibinfo {author}
  {\bibfnamefont{A.}~\bibnamefont{Palazzo}},\ and\ \bibinfo {author}
  {\bibfnamefont{A.~M.}\ \bibnamefont{Rotunno}},\ }%
  \bibfield{journal}{%
  \Doi{10.1016/j.physletb.2005.07.064}{\bibinfo {journal} {Phys. Lett.}}\ }%
  \textbf{\bibinfo {volume} {B623}},\ \bibinfo {pages} {80} (\bibinfo {year}
  {2005}),\ \Eprint{http://arxiv.org/abs/hep-ph/0505081}{arXiv:hep-ph/0505081
  [hep-ph]}%
  \bibAnnoteFile{NoStop}{Fogli:2005qa}%
%%CITATION = HEP-PH/0505081;%%
\bibitem{Lisi:2004jw}%
  \BibitemOpen
  \bibfield{author}{%
  \bibinfo {author} {\bibfnamefont{E.}~\bibnamefont{Lisi}}, \bibinfo {author}
  {\bibfnamefont{A.}~\bibnamefont{Palazzo}},\ and\ \bibinfo {author}
  {\bibfnamefont{A.~M.}\ \bibnamefont{Rotunno}},\ }%
  \bibfield{journal}{%
  \Doi{10.1016/j.astropartphys.2004.03.005}{\bibinfo {journal} {Astropart.
  Phys.}}\ }%
  \textbf{\bibinfo {volume} {21}},\ \bibinfo {pages} {511} (\bibinfo {year}
  {2004}),\ \Eprint{http://arxiv.org/abs/hep-ph/0403036}{arXiv:hep-ph/0403036
  [hep-ph]}%
  \bibAnnoteFile{NoStop}{Lisi:2004jw}%
%%CITATION = HEP-PH/0403036;%%
\end{thebibliography}%

\end{document}